\documentstyle[12pt]{article}
\catcode`\@=11
\ifcase\@ptsize
 \font\tenmsy=msbm10
 \font\sevenmsy=msbm7
 \font\fivemsy=msbm5

\or
 \font\tenmsy=msbm10 scaled \magstephalf
 \font\sevenmsy=msbm8
 \font\fivemsy=msbm6

\or
 \font\tenmsy=msbm10 scaled \magstep1
 \font\sevenmsy=msbm8
 \font\fivemsy=msbm6

\fi
\newfam\msyfam
\textfont\msyfam=\tenmsy  \scriptfont\msyfam=\sevenmsy
  \scriptscriptfont\msyfam=\fivemsy
\def\Bbb{\ifmmode\let\next\Bbb@\else
 \def\next{\errmessage{Use \string\Bbb\space
    only in math mode}}\fi\next}
\def\Bbb@#1{{\Bbb@@{#1}}}
\def\Bbb@@#1{\fam\msyfam#1}
\def\tenrm{}
\def\sevrm{}
\newtheorem{theorem}{Theorem}[section]
\renewcommand{\theequation}{\thesection.\arabic{equation}}

\newcommand{\sect}[1]{\noindent{\hspace{0cm}\large \sl#1}
        \setcounter{equation}{0}\newline}
\newcommand{\subsec}[1]{\noindent{\hspace{0cm}\sl#1}\newline}
\renewcommand{\title}[1]{\bf #1\bigskip\medskip\\}
\renewcommand{\author}[1]{\large #1\\ \smallskip}
\newcommand{\address}[1]{{\normalsize\it #1\\}\bigskip}
\newcommand{\hs}[1]{\hspace*{#1cm}}
\newcommand{\vs}[1]{\vspace*{#1cm}}
\newcommand{\smat}[1]{\mbox{\small $\pmatrix{#1}$}}
\def\and{\;\;{\rm and}\;\;}
\def\disp{\displaystyle}
\def\be{\begin{eqnarray}}
\def\ee{\end{eqnarray}}
\def\no{\nonumber}
\def\half{\mbox{$\textstyle {1 \over 2}$}}
\def\h{\hspace*{0.5cm}}
\def\({\biggl(}
\def\){\biggr)}
\def\l({\left(}
\def\r){\right)}
\def\[{\mbox{$\biggl\lfloor$}}
\def\]{\mbox{$\biggr\rfloor$}}
\def\ZZ{{\Bbb Z}}
\def\CC{{\Bbb C}}
\def\ol{\overline}
\def\talf{\mbox{\small$3\over 2$}}
\def\phalf{\mbox{\small$1\over 2$$p$}}
\def\qhalf{\mbox{\small$1\over 2$$q$}}

\def\T{\mbox{\boldmath { $T$}}}
\def\V{\mbox{\boldmath { $V$}}}
\def\t{\mbox{\boldmath {$t$}}}
\def\scr{\scriptsize}
\def\lam{\lambda}
\def\disp{\displaystyle}
\def\I{\mbox{\boldmath $I$}}

\def\sn{\mbox{\hspace{1.7pt}sn}}

\def\Kp#1#2#3#4{K_{+}\biggl(\matrix{#2\vs{-0.3}&\cr&
  \hs{-0.3}#1\vs{-0.3}\cr#3&}\!\!\biggm|\!\mbox{$#4$}\biggr)}
\def\Km#1#2#3#4{K_{-}\biggl(\!\matrix{&#2\vs{-0.3}\cr\!\!
  #1\hs{-0.3}\vs{-0.3}&\cr&#3}\!\!\biggm|\!\mbox{$#4$}\biggr)}
\def\wt#1#2#3#4#5#6{#1\!\!\mbox{
 $\left(\matrix{#5&#4\cr#2&#3\cr}\biggm|\mbox{$#6$}\right)$}}

\def\AP#1#2#3{\newblock{\sl Ann. Phys.} {\bf #1} (#2) #3}

\def\LMP#1#2#3{\newblock{\sl Lett. Math. Phys.} {\bf #1} (#2) #3}
\def\JSP#1#2#3{\newblock{\sl J. Stat. Phys.} {\bf #1} (#2) #3}
\def\JPA#1#2#3{\newblock{\sl J. Phys.} {\bf A#1} (#2) #3}

\def\IJMP#1#2#3{\newblock{\sl Int. J. Mod. Phys.} {\bf #1} (#2) #3}

\def\NPB#1#2#3{\newblock{\sl Nucl. Phys.} {\bf B#1} (#2) #3}
\def\PLA#1#2#3{\newblock{\sl Phys. Lett.} {\bf A#1} (#2) #3}
\def\PLB#1#2#3{\newblock{\sl Phys. Lett.} {\bf B#1} (#2) #3}
\def\PRL#1#2#3{\newblock{\sl Phys. Rev. Lett.} {\bf#1} (#2) #3}

\def\LMP#1#2#3{\newblock{\sl Lett. Math. Phys.} {\bf#1} (#2) #3}
\def\PA#1#2#3{\newblock{\sl Physica A } {\bf#1} (#2) #3}
\def\SPJ#1#2#3{\newblock{\sl Sov. Phys. JETP. } {\bf#1} (#2) #3}
\begin{document}

\begin{center}
\title{ROW TRANSFER MATRIX FUNCTIONAL RELATIONS FOR
       BAXTER'S EIGHT-VERTEX AND SIX-VERTEX MODELS WITH
      OPEN BOUNDARIES VIA MORE GENERAL REFLECTION MATRICES}
\author{Yu-kui Zhou\footnotemark[1]$^,$\footnotemark[2]
                \footnotetext[1]{Email: {\sl zhouy@maths.anu.edu.au}}
      \footnotetext[2]{ On leave of absence from {\sl Institute of
         Modern Physics,   Northwest University, Xian 710069, China}} }
\address{Mathematics Department, The Australian National University,\\
          Canberra, ACT 0200, Australia }

\end{center}
\begin{abstract}
The functional relations of the transfer matrices of
fusion hierachies for six- and eight-vertex models with
open boundary conditions have been presented in this paper.
We have shown the su($2$) fusion rule for the models with
more general reflection boundary conditions, which are represented by
off-diagonal reflection matrices. Also we have discussed some
physics properties which are determined by the functional relations.
Finally the intertwining relation between the reflection $K$ matrices
for the vertex and SOS models is discussed.

\end{abstract}

\bigskip
to appear in Nucl. Phys. B.
\bigskip

Great progress  has been made on the study of integrable
models in statistical mechanics and quantum field theory. It is very
clear that the important methematical structure that ensures the
exactly solvablity of these models is governed by
  the Yang-Baxter equations if the
models sit on square lattices with periodic boundary conditions.
Recently there has been interest in studying integrable systems with
open boundary conditions. The open boundary conditions are described
by boundary reflection matrices satisfying reflection equations
\cite{Cherednik,Sklyanin} (boundary Yang-Baxter equations),
which ensures the exactly solvablity of
the models with the open boundary conditions together with the
Yang-Baxter equations. The boundary reflection matrices have been
found for many integrable systems, in particular for the six-vertex
and eight-vertex models in \cite{VeGo:93,HoYu:93, InKo:94}
(also see \cite{VeGo:93b,HSFY} for related works).
The eigenvalues of the transfer matrices have been solved for the
six-vertex model or $A_1^{(1)}$ invariant chain
\cite{Sklyanin,ABBBQ:87,MNR:90} (also see \cite{JKKKM} for
related work),
Izergin-Korepin vertex model or $A_2^{(2)}$ invariant chain
\cite{MeNe:92b,YuBa:94},
$U_q(spl(2,1))$-invariant t-J model\cite{FoKa:93,Ruiz94},
$A_n^{(1)}$ invariant chain \cite{VeGo:93b}
and $A_{2n}^{(2)}$ invariant chain \cite{AMN94}.
These exact solutions are constructed by the Bethe ansatz only for
diagonal reflection matrices. The integrable systems with non-diagonal
reflection matrices are more difficult to solve and not
many results have been obtained for such systems.

The fusion procedure has been shown very useful in studying
two-dimensional integrable models. Similar to the method to fuse the
$R$ matrix of the Yang-Baxter equation \cite{KRS:81},
the fusion procedure for the reflection matrix $K$ of the reflection
equation has been presented in \cite{MeNe:92}. The fused  $R$- and
$K$-matrices generate some new integrable models with the open
boundaries based on the elementary model.  The corresponding
fused transfer matrices of fusion hierarchies are related through the
functional relations, which can be shown by the fusion.
In \cite{Zhou:94} the functional relations of transfer
matrices for the six-vertex model with diagonal reflection matrices
have been shown and thus the finite size corrections to the
transfer matrices have been obtained by solving the
functional relations. The functional relations for the six-vertex model
 and eight-vertex model
with non-diagonal reflection matrices have not been found even though
the fusion procedure was described generally before.

In this paper we study the six-vertex model and eight-vertex model with
the non-diagonal reflection matrices found in \cite{VeGo:93,HoYu:93}.
We find the functional relations among the eigenvalues of the transfer
matrices for whole fusion hierarchies of the models. This sheds light on
the solution of the models with boundaries via the most general
non-diagonal reflection matrices. In order to show
the derivation of the functional relations of the six-vertex model and
eight-vertex model with  non-diagonal reflection matrices clearly we
carry out the fusion of the models. Then the intertwining relation
between vertex and SOS models for the boundary reflection $K$ matrices
is also discussed and the reflection equations for the SOS models have
been explained. From the functional relations we derive the crossing
unitary conditions of bulk and surface free energies. For the six-vertex
model the crossing unitary condition has been argued in terms of
quantum field theory in \cite{GhZa:94} and however it is not very clear
for the ($off$-$critical$) eight-vertex model. Our crossing unitary
conditions of bulk and surface free energies must be consistent with
the argument given in \cite{GhZa:94} for the six-vertex model.
We believe
the crossing unitary conditions for the off-critical model is a new
result. Specially, some surface critical behaviour can be studied by
solving the crossing unitary conditions. Also we present the
eigen-spectra of the transfer matrix of the eight-vertex model with
the diagonal reflection matrices, which solves the functional relations.
The eigen-spectra has not been found before.

In section~1 we describe the fusion of the $R$ and $K$ matrices
obviously.
Then the functional equations are presented in section~2. For clarity
all proofs are given in Appendices. In section~3 we present the
functional equations for the six-vertex model and eight-vertex model
with some known boundary reflection matrices. In section~4 the
intertwining relation between the vertex boundary
$K$ matrix and the SOS
face boundary $K$ matrix is discussed. Finally we conclude with a brief
discussion at the crossing unitary conditions for the surface free
energies and the eigen-spectra of the transfer matrix and discuss some
physics consequence.

\sect{\bf 1. Some expressions for fusion}
\renewcommand{\theequation}{1.\arabic{equation}}
Let us begin by reviewing the fusion procedure for a given $R$ matrix
obeying the Yang-Baxter equation
\be
R^{12}(u)R^{13}(u+v)R^{23}(v)&=&R^{23}(v)R^{13}(u+v)R^{12}(u)\;
\label{YBE}
\ee
where $u,v$ are the spectral parameters.
Suppose that $R$-matrix acting on $\CC^2\times\CC^2$ satisfies
the following properties, the $PT$ symmetry
\be
P_{12}R^{12}(u)P_{12}=\stackrel{\!\!\!\!\!\!\!t_1,t_2}{ R^{12}(u)}
   =R^{12}(u)\;\ee
where $P_{12}$ is the permutation matrix, which can be represented by
\be
R^{12}(0)=[\rho(0)]^{1/2}P_{12}\;,
\ee
the unitary condition and the crossing unitary condition
\be
R^{12}(u)R^{12}(-u)&=& \rho(u) \label{unity} \\
 \stackrel{\!\!t_1}{ R^{12}}(u)
 \stackrel{\!\!t_1}{R^{12}}(-u-2\lam)&=&
\tilde{\rho}(u) \;
\label{cross-unity}
\ee
where $t_1$ denotes the transposition in the first space and
$\lam$ is crossing parameter. These
$\rho(u),\tilde{\rho}(u)$ are some scalar $u$-dependent functions.

Fusion is the idea to build up new solutions of the
Yang-Baxter equation. Define the projector
\be
Y_p^+&=&{1\over p!}(P^{1,p}+\cdots+P^{p-1,p}+I)\cdots(P^{1,2}+I)
\ee
the new solutions can be given by \cite{KRS:81}
\be
R_{(p,q)}(u)&=&Y_q^+R_{(p,q)}(u-q\lam+\lam)
 \cdots R_{(p,2)}(u-\lam)R_{(p,1)}(u)Y_q^+ \label{Rpq}\\
R_{(p,j)}(u)&=&Y_p^+R^{1,j}(u)R^{2,j}(u+\lam)\cdots
   R^{p,j}(u+p\lam-\lam)Y_p^+    \;\ee
where $R_{(p,q)}(u)$ acts on $\CC^{p+1}\times \CC^{q+1}$.
These fused $R$ matrices satisfy the fused unitary condition and
the fused crossing unitary condition
\be
R^{12}_{(p,q)}(u)R^{12}_{(q,p)}(-u)&=& \rho_{q,p}(u)Y_pY_q
 \label{funity} \\
 \stackrel{\!\!t_1}{ R^{12}_{(p,q)}}(u)
 \stackrel{\!\!t_2}{R^{12}_{(q,p)}}(-u-2\lam)&=&
\tilde{\rho}_{q,p}(u)Y_pY_q \;
\label{fcross-unity}
\ee
where
\be
\tilde{\rho}_{q,p}(u)&=&\prod_{k=0}^{q-1}\prod_{j=0}^{p-1}
    \tilde{\rho}\(u+(k-j)\lam\)\;.
\ee
Note that the superscripts $1,2$ mean $R^{12}_{(p,q)}(u)\in
\CC^{p+1}\otimes\CC^{q+1}$. We suppress them if there is
no confusion.

The reflection matrices $K_-(u),K_+(u)$
satisfies the reflection equations \cite{Cherednik,Sklyanin}
\be
&&R^{12}(u-v)K^1_-(u)R^{12}(u+v)K^2_-(v) \no \\
&&\h =K^2_-(v)R^{12}(u+v)K^1_-(u)R^{12}(u-v)\; \label{ref-}\\ \hs{-0.7}
&&R^{12}(v\!-\!u)\stackrel{t_1}{K^1_+}\!(u)
   R^{12}(-\!v\!\!-\!\!u\!\!-\!\!2\lam)\stackrel{t_2}{K^2_+}\!(v) \no \\
&&\h =\stackrel{t_2}{K^2_+}\!(v)R^{12}
(-\!v\!\!-\!\!u\!\!-\!\!2\lam)
  \stackrel{t_1}{K^1_+}\!(u)R^{12}(v\!\!-\!\!u)\;.
\label{ref+}
\ee
Similar to the fusion of the $R$ matrix the reflection matrices
$K_-(u),K_+(u)$ acting on space $\CC^2$ can be fused to give
new solutions of the reflection
equations. Following \cite{MeNe:92} the new solutions of the
reflection matrices
can be expressed by
\be
\hs{-1}K^{(q)}_-(u)&=&\rho(u|q)\;Y_q\;[K^{q}_-(u)]
      [R^{q,q-1}(2u+\lam)K^{q-1}_-(u+\lam)]\no \\
 &&\h [R^{q,q-2}(2u+2\lam) R^{q-1,q-2}(2u+3\lam)
                            K^{q-2}_-(u+2\lam)]\cdots\no \\
 &&\h [R^{q,1}(2u+q\lam-\lam)R^{q-1,1}(2u+q\lam)\cdots\no \\
 && \h\h  R^{2,1}(2u+2q\lam-3\lam)K^1_-(u+q\lam-\lam)]\;
 Y_q \label{fK-}\\ \hs{-1}K^{(q)}_+(u)&= &Y_q\;[K^q_+(u)
  \stackrel{t}{R}\!\!^{q-1,q}(-2u-3\lam)\cdots
 \stackrel{t}{R}\!\!^{2,q}(-2u-q\lam)
                  \stackrel{t}{R}\!\!^{1,q}(-2u-q\lam-\lam)]
 \cdots\no \\
 &&\h [K^3_+(u+q\lam-3\lam)\stackrel{t}{R}\!\!^{2,3}(-2u-2q\lam+3\lam)
                  \stackrel{t}{R}\!\!^{1,3}(-2u-2q\lam+2\lam)] \no \\
 &&\h  [K^2_+(u+q\lam-2\lam)\stackrel{t}{R}\!\!^{1,2}(-2u-2q\lam+\lam)]
  [K^1_+(u+q\lam-\lam)] Y_q  \label{fK+}
\ee
where $K^{(q)}_-(u),K^{(q)}_+(u)$ act on $\CC^{q+1}$,
$\stackrel{t}{R}\!\!^{i,j}=
\stackrel{t_it_j}{R}\!\!^{i,j}$ and
\be
\rho(u|q+1)&=&\rho(u|q)/\tilde{\rho}_{q,1}(2u+q\lam) \no\\
\rho(u|1 )&=&1\;, \h q=1,2,\cdots\;.
\ee
The fused reflection equations become
\be
&&R_{(p,q)}(u-v)K^{(p)}_-(u)R_{(q,p)}(u+v-\lam+p\lam)K^{(q)}_-(v)
 \no \\&&\h =\;K^{(q)}_-(v)R_{(q,p)}(u+v-\lam+p\lam)K^{(p)}_-(u)
    R_{(p,q)}(u-v)\; \label{fref-}\\ \hs{-0.7}
&&{\;{R}_{(q,p)}}(v-u)\stackrel{t_1}{K^{(p)}_+}\!(u)
 {\;{R}_{(p,q)}}(-\!v-u-\lam-p\lam)
       \stackrel{\!\!\!t_2}{K^{(q)}_+}\!(v)\no \\
&&\h =\;\stackrel{t_2}{K^{(q)}_+}\!(v){\stackrel{}{R}_{(p,q)}}
(-\!v-u-\lam-p\lam)\stackrel{t_1}{K^{(p)}_+}\!(u)
  {\stackrel{}{R}_{(q,p)}}(v-u)\;.
\label{fref+}
\ee
There is an automorphism between the relation for $K_+^{(p)}$ and
$K_-^{(p)}$ \footnote{some arbitrary parameters in $K_\pm^{(p)}$
are interchanged, e.g. $(\xi_+,\mu_+,\nu_+)
  \leftrightarrow(\xi_-,\mu_-,\nu_-)$
for the six vertex model (\ref{K-six})},
\be
K_+^{(p)}(u)/\rho(u|p-1)&=&
   \stackrel{\!\!\!t}{K}_-\hs{-0.4}{}^{(p)}(-u-p\lam) \\
K_+^{(q)}(u)/\rho(u|q-1)&=&
   \stackrel{\!\!\!t}{K}_-\hs{-0.4}{}^{(q)}(-u-q\lam)\;
\ee
and the automorphism for $R$-matrices
\be
R_{(q,p)}(u)=R_{(p,q)}(u+q\lam-p\lam)
\ee
which can be seen directly from the fusion procedure.

The fused Yang-Baxter equation and the fused reflection equation for
$R$ and $K$ matrices guarantee  the following commuting families
\be
\left[\T^{(p,q)}(u)\; ,\; \T^{(p,b)}(v)\;\right]=0\; \label{TT}
\ee
where
\be
&&\T^{(p,q)}(u)={\rm tr}\(\;
 \stackrel{t}{K}_+\!\!\!\!^{(q)}(u)\mbox{\boldmath $U$}_{(p,q)}(u)
  K_-^{(q)}(u)\tilde{\mbox{\boldmath $U$}}_{(p,q)}
  (u+q\lam-\lam) \) \label{fT-o} \\
&&\hs{-1.5}\mbox{\boldmath$U$}_{(p,q)}(u)=
        R^{c,1}_{(q,p)}(u+\phalf\lam-\half\lam)
      R^{c,2}_{(q,p)}(u+\phalf\lam-\half\lam)
       \cdots,R^{c,N}_{(q,p)}(u+\phalf\lam-\half\lam)\; \label{U} \\
&&\hs{-1.5}\tilde{\mbox{\boldmath $U$}}_{(p,q)}(u)=
   R^{N,c}_{(p,q)}(u-\phalf\lam+\half\lam)\cdots
    R^{2,c}_{(p,q)}(u-\phalf\lam+\half\lam)\;
    R^{1,c}_{(p,q)}(u-\phalf\lam+\half\lam)\;   \label{U1}
\ee
and $q,p,b=1,2,\cdots$. The proof of (\ref{TT}) can be done
similarly to the study of the unfused six-vertex model given by
Sklyanin \cite{Sklyanin} and is briefly described in Appendix~A.
using graph representation.

\sect{\bf 2. Functional equations for general reflection matrices}
\renewcommand{\theequation}{2.\arabic{equation}}
The last section has shown that  fused models can be built up by fusion
of the elementary $R$ and $K$ matrices. The fused models are integrable
systems because we have the commuting families (\ref{TT}). In fact, the
these fused transfer matrices are related each other by the groups of
functional relations, which can be constructed by fusion. For the
periodic boundary condition it was already well known that the
transfer matrix
\begin{equation}
\T^{(p,q)}(u)\;=\;{\rm tr}\;\mbox{\boldmath$U$}_{(p,q)}(u)
\end{equation}
commutes
\be
\left[\T^{(p,q)}(z)\; ,\; \T^{(p,q')}(y)\;\right]=0\; \label{pTT}
\ee
and the functional relations
\be
&&\T^{(p,q)}(u)\T^{(p,1)}(u+q\lam)=
 \T^{(p,q+1)}(u)+f^p_{q-1}\T^{(p,q-1)}(u)\;
\ee
can be constructed \cite{KiRe:87},
where $\T^{(p,0)}(u)\;=\;\I\in\CC^{p+1}$ and the
$u$-dependent function $f^p_q$  is generated from the
antisymmetric fusion of the model. These relations,
in fact, are the su($2$) fusion rule and show the implication of the
eigenvalues of transfer matrices of all fusion hierarchies. They have
been applied successfully to derive the eigenvalues and the relevant
Bethe ansatz equations of the fused or the elementary transfer
matrices \cite{KiRe:87,BaRe:89}. Also they could be solved directly
in the thermodynamic limit to yield the finite size corrections to the
eigenvalues of the transfer matrices
\cite{KlPe:92,KNS:94,ZhPe:95,Zhou:94}.

For the models with open reflection boundary condition it has been
noticed that the functional relations can be applied to obtain the
eigenvalues of fused transfer matrices \cite{MNR:90,MeNe:92} and
to find the finite size corrections to the
eigenvalues of transfer matrices
\cite{Zhou:94}. These studies have concerned only  models
with diagonal reflections. For the off-diagonal reflection
boundary solvable models we still expect the su($2$) fusion rule. This
is shown in this section.

Let us first prepare some notations as follows.
\be \hs{-0.5}
 \mbox{\boldmath $\phi$}_+^p(u)&:=&
      \prod_{i=1}^{N}[Y_2^-R_{(1,p)}^{1,\;i}(u+\phalf\lam-\half\lam)
           R_{(1,p)}^{2,\;i}(u+\phalf\lam+\half\lam)Y_2^-]
    \in\otimes_{i=1}^{N}\CC^{p+1}\no \\
\mbox{\boldmath $\phi$}_-^p(u)&:=&
    \prod_{i=1}^{N}[Y_2^-R_{(p,1)}^{i,\;1}(u-\phalf\lam+\half\lam)
           R_{(p,1)}^{i,\;2}(u-\phalf\lam+\talf\lam)Y_2^-]
    \;\in\otimes_{i=1}^{N}\CC^{p+1} \no \\
\I_+(u|q)&:=&Y_2^-R^{1,q-1}_{(1,q-1)}(
   q\lam-4\lam-2u)R^{2,q-1}_{(1,q-1)}(q\lam-3\lam-2u)Y_2^-
 \in\CC^{q} \label{part} \\
\I_-(u|q)&:=& \disp{Y_2^-R^{q-1,1}_{(q-1,1)}
  (2u-q\lam+\lam)R^{q-1,2}_{(q-1,1)}(2u-q\lam+2\lam)Y_2^-
 \over\tilde{\rho}_{q,1}^{}(2\!u\!\!-\!\!q\!\lam\!\!+\!\!2\!\lam)
 \tilde{\rho}_{q\!-\!1,1}^{}(2\!u\!\!-\!\!q\!\lam\!\!+\!\!\lam)}
 \;\in\CC^{q}  \no
\ee
where $Y_2^-=\half(1-P^{1,2})$ is an antisymmetric projector.
The boundary dependent notations are defined by
\be
\omega_+(u)&:=&Y_2^-\stackrel{t}{K}_+\!\!\!\!^{1}
  (u+\lam)R^{1,2}(-2u-3\lam)
   \stackrel{t}{K}_+\!\!\!\!^{2}(u)Y_2^- \no \\
\omega_-(u)&:=&Y_2^-K_-^{1}(u)R^{1,2}(2u+\lam)K_-^{2}(u+\lam)Y_2^-
\label{anti-K}\ee
The product of all of these notations is expressed by
\be
\mbox{\boldmath $f$}^p(u)=
 \omega_-(u)\omega_+(u)\(\mbox{\boldmath $\phi$}_+^p(u)
   \mbox{\boldmath $\phi$}_-^p(u)\) \otimes\(\I_+(u|q)\I_-(u|q)\)
  \;\;\in\otimes_{i=1}^{N}\CC^{p+1}\otimes\CC^{q} \; , \label{bf}
\ee
which is clearly a $q$-independent function  if we have the relation
\be
\I_+(u|q)\I_-(u|q)=\I\; \tilde{\rho}_{1,1}^{-1}(2u+\lam)\h \;,\label{II}
\ee
where the identity matrix depends on the fusion
level $q$ or $\I\in\CC^{q}$. The above relation (\ref{II})
is just right for the six- and eight-vertex
models, which will be seen in the next section.
Moreover \mbox{\boldmath $\phi$}
and \mbox{\boldmath $I$} are the diagonal matrices and so is
$\mbox{\boldmath $f$}^p(u)$, which is written in the form
\be
\mbox{\boldmath $f$}^p(u)&=& \I \cdot f^p(u)\; \no \\
f^p(u)&=&\omega_-(u)\omega_+(u)\phi_+^p(u)\phi_-^p(u)I_+(u|q)I_-(u|q)
\label{f}\ee
and $\I\in\otimes_{i=1}^{N}\CC^{p+1}\otimes\CC^{q}$.
Now we can express the
functional relations for  the six- and
eight-vertex models  with open reflection boundary
conditions in the following theorem.
\begin{theorem}[$su(2)$ Fusion Hierarchy]\label{Ther-1}
Let us define
$$  f^p_{q}\;=\;f^p(u+q\lam)\; ,\h\h\T^{(p,0)}\;=\;\I
    \h \in\disp{\otimes_{i=1}^N\CC^{p+1}}$$
with the transfer matrices
$$T^{(q)}_k\;=\;T^{(p,q)}(u+k\lam)$$
given by (\ref{fT-o}). Then the su($2$) fusion hierarchy follows as
\be
\h \T^{(q)}_0\T^{(1)}_q= \T^{(q+1)}_0  + f^p_{q-1}\T^{(q-1)}_0\;
  \h\h q=1,2,\cdots
\label{Func-T}  \ee
if the reflection $K_-,K_+$ matrices satisfy the reflection
equations (\ref{ref-})-(\ref{ref+}).
\end{theorem}
The theorem is proved in Appendix~B. Then following the standard
procedure given by \cite{KlPe:92,KuNa} by introducing the inversion
identity hierarchy (or $y$-system)
\be
\t^{(0)}_0&=&0 \;     \\
\t^{(q)}_0&=&\disp{\T^{(q+1)}_0\T^{(q-1)}_1/
   \disp\prod_{k=0}^{q-1}f^p_k }\; ,
\label{def-t}\ee
the functional equations for $\t$ can be derived, which are
called thermodynamic Bethe ansatz (TBA) equations
in \cite{KlPe:92,Zhou:94}. To see this let us consider the triple
transfer matrices
$$ \T^{(p,q)}_0(\T^{(p,q-1)}_1\T^{(p,1)}_q)=
(\T^{(p,q)}_0\T^{(p,1)}_q)\T^{(p,q-1)}_1\; .$$
Inserting the functional relations (\ref{Func-T}) into the terms in
parentheses this equation gives new functional
equations,
\be
\T^{(q)}_0\T^{(q)}_1&=&
 \I\prod_{k=0}^{q-1}f^p_k  + \T^{(q+1)}_0\T^{(q-1)}_1\;.
\ee
This leads to the following theorem.
\begin{theorem}[$su(2)$ TBA]\label{Ther-2}
The inversion identity hierarchy $\t^q$ satisfies the following
$su(2)$ TBA equations
\be
\t^{(q)}_0\t^{(q)}_1=(\I+\t^{(q+1)}_0)(\I+\t^{(q-1)}_1)\;\label{Func-t}
\ee
for any reflection $K_-,K_+$ matrices satisfying the reflection
equations (\ref{ref-})-(\ref{ref+}),
 where $\I\in\otimes_{i=1}^{N}\CC^{p+1}$.
\end{theorem}

\sect{\bf 3. Examples}
\renewcommand{\theequation}{3.\arabic{equation}}
We have shown that the functional relations hold for the  six-
and eight-vertex models with any $K$ reflection matrices satisfying
the reflection equations (\ref{ref-})-(\ref{ref+}). The $su(2)$
fusion rule depends on the boundaries though the function
$f^p(u)$ defined in (\ref{f}). In this section we calculate
explicitly the function for the six- and eight-vertex models.
Here we use the general $K$ matrices  presented in
\cite{VeGo:93} and \cite{HoYu:93}.

\subsec{\bf 3.1 Six vertex model}
The six vertex model is described by the following $R$ matrix,
\be
&&\hs{1}R(u)=\smat{a(u)&0&0&0\cr
           0&b(u)&c(u)&0\cr
           0&c(u)&b(u)&0\cr
           0&0&0&a(u)} \;     \vs{0.3}   \label{sixR} \\
&&\hs{-0.8}a(u)=\sin(u+\lam)\;, \hs{0.8}b(u)=\sin(u)\;,\hs{0.8}
c(u)=\sin(\lam) \;      \no
\ee
and the boundary reflection matrices are \cite{VeGo:93}
\be
K_+(u)&=& \smat{\sin(\xi_+-u-\lam)&\nu_+\sin(2u+2\lam)\cr
              \mu_+\sin(2u+2\lam)&\sin(\xi_++u+\lam)}       \\
K_-(u)&=&\smat{\sin(\xi_-+u)&\mu_-\sin(2u)\cr
              \nu_-\sin(2u)&\sin(\xi_--u)}  \label{K-six}
\ee
where $\xi_\pm,\mu_\pm,\nu_\pm$ are arbitrary parameters.

The unitary conditions can be checked by inserting the $R$ matrix
(\ref{sixR}) into (\ref{unity}) and (\ref{cross-unity}) and we
obtain
\be
\rho(u)&=& \sin^2(u)-\sin^2(\lam)\; ; \h
\tilde{\rho}(u)\;=\;\sin^2(\lam)-\sin^2(u+\lam)=-b(u)a(u+\lam)
\ee
The quantities in (\ref{part}) are given by calculation to be
\be
&&\omega_+(u)=-\sin(2u\!+\!4\lam)\(\sin(\xi_+\!+\!u\!+\!\lam)
  \sin(\xi_+\!-\!u\!-\!\lam)-\mu_+\nu_+\sin^2(2u\!+\!2\lam)\)   \no\\
&&\omega_-(u)=\sin(2u)\(\sin(\xi_-\!+\!u\!+\!\lam)\sin(\xi_-
 \!-\!u\!-\!\lam)-\mu_-\nu_-\sin^2(2u\!+\!2\lam)\) \label{sixomega} \\
&&\phi^p_+(u)=\;\prod_{k=0}^{p-1}[-\tilde{\rho}_{1,1}
         (u-k\lam+\phalf\lam-\half\lam)]^N                \no \\
&&\phi^p_-(u)=\;\prod_{k=0}^{p-1}[-\tilde{\rho}_{1,1}
          (u+k\lam-\phalf\lam+\half\lam)]^N         \\
&&I_+(u|q)=\;\prod_{k=1}^{q-1}
     [-\tilde{\rho}_{1,1}(2u+\lam-k\lam)]      \no   \\
&&I_-(u|q)=-\;\prod_{k=1}^{q}
    [-\tilde{\rho}_{1,1}(2u+2\lam-k\lam)]^{-1} \;.    \label{sixI}
\ee
It is obvious that
the relation (\ref{II}) is satisfied.

The fusion procedure for the $R$ matrix brings
the extra zeros to the fused
$R$ matrices. The number of the zeros are only dependent on the fusion
level $p,q$. Some zeros of  $R_{(p,q)}(u)$ can be removed from
$\mbox{\boldmath $U$}_{(p,q)}(u)$ and
$\tilde{\mbox{\boldmath $U$}}_{(p,q)}(u+q\lam)$
in (\ref{U}) and (\ref{U1}) by the  replacements
\be
&&R_{(q,p)}(u)\to R_{(q,p)}(u)/\prod_{j=0}^{q-1}\prod_{k=0}^{p-2}
[b(u+j\lam-k\lam)]\;    \no \\
&&R_{(p,q)}(u)\to R_{(p,q)}(u)/\prod_{k=0}^{q-1}\prod_{j=1}^{p-1}
[b(u+j\lam-k\lam)]\;
\ee
The replacement only changes $\phi_\pm^p(u)$ to a simpler form,
\be
\phi_+^p(u)= \phi_-^p(u)=[b(u-\phalf\lam+\half\lam)
   a(u+\phalf\lam+\half\lam)]^N \label{sixphi}
\ee
Inserting all of the quantities given in
(\ref{sixomega})-(\ref{sixphi})
into the definition (\ref{f}) we have
\be
f^p(u)=\omega_+(u)\omega_-(u){[b(u-\phalf\lam+\half\lam)
    a(u+\phalf\lam+\half\lam)]^{2N}\over
 b(2u+\lam)a(2u+2\lam)}\label{sixf}
\ee
For the simple case of  diagonal $K$ matrices the fusion rule
(\ref{Func-T}) with (\ref{sixf}) coincides with that obtained
in \cite{Zhou:94}.

\subsec{\bf 3.2 Eight-vertex model}
The eight-vertex model is the generalization of the six-vertex model
\cite{Baxter}. The $R$ matrix for the eight-vertex model is given by
\be
&&\hs{1.1}R(u)=\smat{a(u)&0&0&d(u)\cr
           0&b(u)&c(u)&0\cr
           0&c(u)&b(u)&0\cr
           d(u)&0&0&a(u)} \;  \vs{0.3} \label{R-8}
\ee
These non-zero elements are given by
\be
a(u)=H(u+\lam)\Theta(u)\Theta(\lam)\;, &&
b(u)=\Theta(u+\lam)H(u)\Theta(\lam)\;,     \no \\
c(u)=\Theta(u+\lam)\Theta(u)H(\lam)\;, &&
d(u)=H(u+\lam)H(u)H(\lam)\;. \label{abcd}
\ee
where $H(u),\Theta(u)$ are the $theta$ functions.
The relevant boundary reflection matrices are given by
\cite{HoYu:93}
\be
&&K_-(u)\;=\;\smat{H(\xi_-+u)
 \Theta(\xi_--u)&\mu_-\hspace{1.7pt}
     k^{1/2}\sn(2u)\sn^2(u)/f(u;\xi_-)\cr
  -{\mu}_-\hspace{1.7pt}k^{-1/2}
 \sn(2u)/f(u;\xi_-)&\Theta(\xi_-+u)H(\xi_--u)}
\label{K-eight}    \\
&&K_+(u)=     \\
&&\smat{H(\xi_+-u-\lam)\Theta(\xi_++u+\lam)&
  \mu_+\hspace{1.7pt}k^{-1/2}\sn(2u+2\lam)/f(u+\lam;\xi_+)\cr
  -{\mu}_+\hspace{1.7pt}k^{1/2}\sn(2u+2\lam)
   \sn^2(u+\lam)/f(u+\lam;\xi_+)&
    \Theta(\xi_+-u-\lam)H(\xi_++u+\lam)}  \no
\ee
with
 $$f(u;\xi)=[1-k^2\sn^2(u)\sn^2(\xi)]/[\Theta(\xi+u)\Theta(\xi-u)]\;,$$
and
$$ \sn (u)=k^{-1/2}H(u)/\Theta(u) $$
where $\xi_\pm,\mu_\pm$ are arbitrary parameters.
The scalar functions in the unitary relations (\ref{unity}) and
(\ref{cross-unity}) for the eight vertex models are
\be
\rho(u)&=&h(\lam+u)h(\lam-u) \\
\tilde{\rho}(u)&=&h(u)h(u+2\lam)
   \label{eightrho}\\
&&h(u)\;=\;H(u)\Theta(u)\Theta(0)
\ee
The expressions of the quantities in (\ref{sixI}) with (\ref{eightrho})
are still correct except
the $\omega_\pm(u)$, or
\be
&&\omega_-(u)=h(\xi_-+u+\lam)
  h(\xi_--u-\lam)H(2u)\Theta(2u+2\lam)/\Theta(0) \\
&& \h +  K_-(u)_{21}K_-(u+\lam)_{21}H(2u+2\lam)H^2(u+\lam)
  \Theta(2u)\Theta(0)/\Theta^2(u) \no \\
&&\omega_+(u)=-h(\xi_+-u-\lam)
 h(\xi_++u+\lam)H(2u+4\lam)\Theta(2u+2\lam)/\Theta(0) \\
&& \h -  K_+(u+\lam)_{12}K_+(u)_{12}H(2u+2\lam)H^2(u+\lam)
  \Theta(2u+4\lam)\Theta(0)/\Theta^2(u+2\lam) \no \\
%&&\omega_+(u)=-2\sn(u+2\lam)\cn(u+2\lam)\dn(u+2\lam)        \no\\
%&& \h\h\h\h \times\sn(\xi_++u+\lam)\sn(\xi_+-u-\lam)
%             {f(u+\lam;\xi_+)^2\over f(u+2\lam;u+\lam)} \no\\
%&&\h +\mu_+^2k^{-1}\sn^2(2u+2\lam)\sn(2u+4\lam)
%        \(\sn(2u+3\lam)\sn(\lam)-\sn^2(u+2\lam)\)   \no  \\
%&&\omega_-(u)=2\sn(u)\cn(u)\dn(u)\sn(\xi_-+u+\lam)\sn(\xi_--u-\lam)
%         {f(u+\lam;\xi_-)^2\over f(u+\lam;u)} \no \\
%&&\h  +\mu_-^2k^{-1}\sn^2(2u+2\lam)\sn(2u)
%       \(\sn(2u+\lam)\sn(\lam)+\sn^2(u)\)\label{eightomega} \\
&&\phi^p_+(u)=\;\prod_{k=0}^{p-1}[-\tilde{\rho}_{1,1}
         (u-k\lam+\phalf\lam-\half\lam)]^N                \no \\
&&\phi^p_-(u)=\;\prod_{k=0}^{p-1}[-\tilde{\rho}_{1,1}
          (u+k\lam-\phalf\lam+\half\lam)]^N         \\
&&I_+(u|q)=\;\prod_{k=1}^{q-1}
     [-\tilde{\rho}_{1,1}(2u+\lam-k\lam)]      \no   \\
&&I_-(u|q)=-\;\prod_{k=1}^{q}
    [-\tilde{\rho}_{1,1}(2u+2\lam-k\lam)]^{-1} \;.    \label{eightI}
\ee
{}From (\ref{eightI}) it can be seen that (\ref{II}) is satisfied.

Similar to the six-vertex model discussed in the last subsection,
removing the zeros generated by fusion from
$\mbox{\boldmath $U$}_{(p,q)}(u)$ and
$\tilde{\mbox{\boldmath $U$}}_{(p,q)}(u+q\lam)$
in (\ref{U}) and (\ref{U1})
\be
&&R_{(q,p)}(u)\to R_{(q,p)}(u)/\prod_{j=0}^{q-1}\prod_{k=0}^{p-2}
[h(u+j\lam-k\lam)]\;    \no \\
&&R_{(p,q)}(u)\to R_{(p,q)}(u)/\prod_{k=0}^{q-1}\prod_{j=1}^{p-1}
[h(u+j\lam-k\lam)]\;
\ee
from the transfer matrix (\ref{fT-o}), the quantities $\phi_\pm^p(u)$
become
\be
\phi_+^p(u)&=&[h(u-\phalf\lam+\half\lam)
          h(u+\phalf\lam+\half\lam+\lam)]^N \no \\
\phi_-^p(u)&=&            %[F_p^-(u-\phalf\lam+\half\lam)
  [h(u-\phalf\lam+\half\lam)
   h(u+\phalf\lam+\half\lam+\lam)]^N  \;.\label{eightphi}
\ee
Inserting all of the quantities given in
(\ref{sixomega})-(\ref{sixphi})
into the definition (\ref{f}) we have
\be
f^p(u)=\omega_+(u)\omega_-(u){[h(u-\phalf\lam+\half\lam)
    h(u+\phalf\lam+\half\lam+\lam)]^{2N}\over h(2u+\lam)h(2u+3\lam)}\;.
\label{eightf}\ee

\sect{\bf 4. SOS models and intertwiner}
\renewcommand{\theequation}{4.\arabic{equation}}
Intertwining relations are the correspondence between
the $R$-matrix (the
Boltzmann weights of the vertex model) and $W$-matrix (the Boltzmann
weights of the relevant face model) \cite{Baxter:73,JMO:87}. Baxter
introduced the intertwining relation for the eight-vertex and the
relevant face model to solve the eigenvalues
of transfer matrices of the
eight-vertex model. The related face model is called the solid-on-solid
(SOS) model. The similar formulation can be generalized to the boundary
$K$ matrices (boundary Boltzmann weights). This however has not been
done before. We show briefly in this section
the correspondence between
the boundary $K$ matrices of the eight-vertex and the relevant SOS
models. For clarity the corresponding graph representations are briefly
given in appendix~C.

At first let us review  Baxter's intertwining relation between
the eight-vertex and the SOS model. Choose arbitrary constants
$s^\pm$ and integers $a,b\in\ZZ$ and set the nonzero
intertwiners by
\be
\varphi_{a,b}(u)=\smat{H(s^\epsilon+a-\epsilon u)\cr
      \Theta(s^\epsilon+a-\epsilon u)}/\sqrt{S(b)}\h
    \mbox{if $\epsilon=b-a=\pm 1\;$} \label{phi}
\ee
where the factors $S(b)$ is given by
\be
S(b)&=&{h(w_0+b\lam)}   %\no \\%{} [u]&=&\theta_1(\pi u/2K,p)\;.
\ee
The $R$ matrix is given by (\ref{R-8}) with (\ref{abcd}).
Then  Baxter's intertwining relation is given by
\be
R(u-v)\;\varphi_{d,c}(u)\otimes\varphi_{c,b}(v)=\sum_{a}
 \varphi_{d,a}(v)\otimes\varphi_{a,b}(u)\;\wt Wabcd{u-v} \;,
\label{RW}\ee
where $\wt Wabcd{u}$ are the face weights of the SOS model
satisfying the star-triangle relation
\be
&&\sum_g\wt Wabgfu\wt {W}fgdev\wt {W}gbcd{v\!-\!u} \no \\
&&\h=\sum_g\wt {W}fage{v\!-\!u}\wt {W}abcgv\wt Wgcdeu \label{STR}
\ee
for any $a,b,c,d,e,f.\;$
The nonzero weights are given by
\be
\wt W{a\pm 1}{a\pm 2}{a\pm 1}a{u}&=& {h(\lam+u)\over h(\lam)}\no\\
\wt W{a\mp 1}{a}{a\pm 1}a{u}&=&\sqrt{S(a+1)S(a-1)
          \over S(a)^2}     {h(u)\over h(\lam)}  \\
\wt W{a\pm 1}{a}{a\pm 1}a{u}&=&{h(w_0+a\lam\mp u)\over h(w_0+a\lam)}\no
\ee
where $a\in\ZZ$ and $w_0=\half(s^++s^-)\lam-K$. These face weights
satisfy the following unitary condition
\be
\sum_{c}\wt Wabcd{u} \wt Wcb{a'}d{-u}=
\rho(u)\delta_{aa'}\; \label{Wunit}
\ee
and the crossing unitary condition
\be
\sum_{c}\wt Wbcda{-u-\lam} \wt Wb{a'}dc{u-\lam}
 {S(a)S(c)\over S(b)S(d)}=\rho(u)\delta_{aa'}\;,
  \label{Wcrossunit} \ee
where the scalar function is given by
\be
\rho(u)=\disp{{h(\lam-u)\over h(\lam)}{h(\lam+u)\over h(\lam)}}\;.
\ee

The intertwining relation for the boundaries is set out by
\be
K_-(u)\varphi_{b,a}(u)&=&\sum_c\varphi_{b,c}(-u)\Km bacu \no \\
 \stackrel{t}{K}_+(u)\varphi_{b,a}(-u-\lam)&=&\sum_c
  {S(c)\over S(b)}\varphi_{b,c}(u+\lam)\Kp bcau
\; ,\label{KK}\ee
where $\Km bacu$ and $\Kp bcau$ are
the reflection matrices ($K$ matrices)
of the SOS model. It is obvious that the face $K$ matrix elements
are nonzero only for $|b-a|=1$ and $|b-c|=1$. The nonzero
elements are found to be
\be
\Km bacu&=&\varphi_{b,c}^{-1}(-u)K_-(u)\varphi_{b,a}(u) \no \\
\Kp bacu&=&\varphi_{b,a}^{-1}(u+\lam)\stackrel{t}{K}_+(u)
    \varphi_{b,c}(-u-\lam){S(b)\over S(a)}
\ee

The intertwiners are dependent on two arbitrary parameters $s^\pm$
and thus $\varphi_{a,a+1}(u)$ and $\varphi_{a,a-1}(u)$ can be treated
as independent vectors. We therefore can define the inverse
vectors of them by
\be
\sum_{b=a\pm 1}\varphi_{a,b}(u)_i\varphi_{a,b}^{-1}(u)_j&=&
  \delta_{i,j} \no \\
\sum_{i=1,2}\varphi_{a,b}^{-1}(u)_i\varphi_{a,c}(u)_i&=&
  \delta_{a,c}\;
\ee
where
\be
\varphi_{a,b}^{-1}(u)=
  S(b)\(\epsilon\Theta(s^{-\epsilon}+a+\epsilon u),\;
      -\epsilon H(s^{-\epsilon}+a+\epsilon u)\)/det_a[\varphi(u)]\h
        \; \label{phi-1}
\ee
with $\epsilon=b-a=\pm 1$  and
$$det_a[\varphi(u)]=
  {2H(x-K)\Theta(x-K)H(y)\Theta(y)\over H(K)\Theta(K)}$$
with $x=\half(s^- +s^+)+a\lam$ and $y=\half(s^- -s^+)+u$.

Applying the intertwining relations (\ref{RW}) and (\ref{KK})
to the reflection equations for the vertex model we are able to obtain
the reflection equations for the SOS model,
\be
&&\sum_{f,g}{\wt Wgcba{u-v}}{\Km gcfu}{\wt Wdfga{u+v}}{\Km dfev}
   \label{WK-}\\
&&\h =\sum_{f,g} {\Km bcfv}{\wt Wgfba{u+v}}{\Km gfeu}{\wt Wdega{u-v}}
     \no \\
&&\hs{-0.5}\sum_{f,g}{\wt Wbcga{v\!-\!u}}{\Kp gcfu}
  {\wt Wdfga{\!\!-\!2\!\lam\!\!-\!\!u\!\!-\!\!v}}{\Kp dfev}
   {S(f)S(a)\over S(d)S(g)} \label{WK+}\\
&& =\sum_{f,g} {\Kp bcfv}
 {\wt Wgfba{\!\!-\!2\!\lam\!\!-\!\!u\!\!-\!\!v}}
   {\Kp gfeu}{\wt Wdega{v\!-\!u}} {S(f)S(a)\over S(b)S(g)}
   \;.  \no
\ee
This SOS analogue of the reflection equations, directly following
from the reflection equations of the vertex model, has not been given
before \cite{comm0}.
Similar to the vertex model, we have the commuting transfer matrix
$\mbox{\boldmath $V$}(u)$ defined by the following elements
\be
\langle\mbox{\boldmath $a$}|{\bf V}(u)|\mbox{\boldmath $b$}\rangle
&=&\sum_{\{c_0,\cdots,c_N\}} \Kp {c_0}{a_0}{b_0}u\prod_{k=0}^{N-1}
 \biggl[ \wt W{b_k}{b_{k+1}}{c_{k+1}}{c_k}{u\!+\!v_k} \no \\
&& \h\h\times\wt W{c_{k+1}}{a_{k+1}}{a_k}{c_k}{u\!-\!v_k}\biggl]
 \Km {c_N}{a_N}{b_N}u  \; ,  \label{V}
\ee
which satisfies
\be
\left[\V(u)\; ,\; \V(v)\;\right]=0\; \label{VV}
\ee
where $\mbox{\boldmath $a$}=\{a_0,a_1,\cdots,a_N\}$,
$\mbox{\boldmath $b$}=\{b_0,b_1,\cdots,b_N\}$ and these $v_k$ are
some arbitrary parameters. The proof of
(\ref{VV}) can be done similarly to the vertex models. Also
the  su($2$) type fusion rule in theorem~(\ref{Ther-1})
still works for the SOS models with open reflection boundaries.

\begin{theorem}[$su(2)$ Fusion Hierarchy]\label{Ther-3}
Let us define that
$$  V^{(q)}_k\;=\;V^{(p,q)}(u+k\lam)\; ,\h\h\V^{(p,0)}\;=\;\I $$
are the fused transfer matrices of the SOS model with the fusion levels
$p$ and $q$.  Then the su($2$) fusion hierarchy follows that
\be
\h \V^{(q)}_0\V^{(1)}_q= \V^{(q+1)}_0  + f^p_{q-1}\V^{(q-1)}_0\;
  \h\h q=1,2,\cdots
\label{func-V}  \ee
if the reflection $K_-,K_+$ matrices satisfy the reflection
equations (\ref{WK-})-(\ref{WK+}).
The function $f^p_{q}\;=\;f^p(u+q\lam)$
is given by the antisymmetric fusion.
\end{theorem}

The proof can be done similarly to the case of vertex models. In
appendix~C we give the proof for the simpler case of $q=1$ and $p=1$.
It is interesting to notice that the
functional relations (\ref{func-V})
of the SOS model with the open boundaries and periodic boundaries
have the same form except the $u$-dependent function $f^p_{q-1}$ being
different.

\sect{\bf 5. Discussion}
In the previous sections we have studied the fusion hierarchies of the
six and eight vertex models with the open boundaries. The functional
relations of  the fused transfer matrices of the models with
the non-diagonal reflection matrices have been obtained. It has been
shown that the functional relations can be solved to obtained the
finite size corrections to transfer matrices for the six-vertex model
with the diagonal reflection matrices \cite{Zhou:94}. It should be
interesting in the near future to solve the fusion hierarchy and
thermodynamic Bethe ansatz equations for the six-vertex  model with
the general non-diagonal reflection matrices by similar techniques.
In this way it should be possible to obtain the central charges
and scaling dimensions in terms of
Rogers dilogarithms and their analytic
continuations and hence completely elucidate the critical behaviour of
the model and their fusion hierarchies.

The free energies and the eigen-spectra of the transfer matrices
of the models with the open boundaries can be determined by
the functional relations. To show this the discussion can be divided
into a number of points.

\noindent{\bf 5.1. Crossing unitary condition for boundaries:}
\renewcommand{\theequation}{5.\arabic{equation}}
The crossing symmetry for boundary $K$ matrices have been expressed in
\cite{GhZa:94}. Using the crossing symmetries
and the unitary conditions
of the boundary $K$ matrices the surface free
 energy $f_s$ can be fixed.
In \cite{GhZa:94} the argument is given  in terms of quantum field
theory. According to this argument it is not very clear how to express
the crossing and unitary relation of the surface free energy for an
off-critical model. However, like the $R$ matrix the crossing symmetry
of the $K$ matrix can be given by the fusion procedure in statistical
mechanics. To see this let us first
consider the crossing symmetry of the $R$ matrix.

The antisymmetric fusion of the $R$ matrix
\be
Y_2^-R^{1,1}(u-\lam)R^{2,1}(u)Y_2^-=
 -\tilde{\rho}_{1,1}(u-\lam)={\rho}_{1,1}(u)
\ee
is equivalent to the crossing unitary condition (\ref{cross-unity}).
This relation leads to the crossing unitary condition
for the bulk free energy
\be
f_b(u)f_b(u+\lam)=[{\rho}_{1,1}(u)]^{2N}
\ee
Similarly, the crossing unitary condition of the $K$ matrices is
related to the antisymmetric fusion of the $K$ matrices
(see (\ref{anti-K})). More precisely the factors $\omega_\pm$ determine
the crossing unitary conditions of the
 $K_\pm$ matrices. Thus the surface
free energies $f_s$ are determined by
\be
f_s(u)f_s(u+\lam)={\omega_-(u-\lam)\omega_+(u-\lam)\over
    h(2u-\lam) h(2u+\lam)} \label{fs}
\ee
The denominator $h(2u-\lam)h(2u+\lam)$ or
$\sin(2u-\lam)\sin(2u+\lam)$
in the function $f^1(u-\lam)$ can be
removed by normalization of the transfer matrix $T^{(1,1)}$.
Thus the functions $f^p(u)$ in (\ref{sixf}) or (\ref{eightf}) contain
the unitary conditions and crossing symmetries of the bulk
$R$ and the boundary $K$ matrices.

The bulk free energy of the eight-vertex model has been obtained by
Baxter \cite{Baxter:71}. By the similar method we can solve the
the (\ref{fs}) to obtain the surface free energy for the off-critical
model. Therefore we should have the surface specific heats and the
related critical exponents. This study does not rely on the exact
diagonalization of the transfer matrix of the model with
the non-diagonal reflection matrices, which has not yet solved.
Thus we have presented a possible method to find the crossing unitary
conditions for solving the surface free energy of an
exact solvable model with open boundary conditions.

\noindent{\bf 5.2. Ansatz of eigen-spectra:}
We have found the functional relations of the transfer matrices for all
fusion hierarchies of the six- and eight-vertex models with the most
general reflection matrices. The su($2$) fusion rule still hold for the
models with any $K$ matrices satisfying the reflection equations.
It is not very clear to extract the exact eigen-spectra of
transfer matrices from the functional relations for the models with
general reflection matrices. However we may solve the eigen-spectra of
transfer matrix from the functional relations for the eight-vertex model
with the diagonal reflection matrices, which, in fact, has not been
given before .

For the special case of  diagonal $K$ matrices, the transfer matrix
$T(u)=T^{(1,1)}(u)$ of the six vertex model has been diagonalized using
algebraic Bethe ansatz \cite{Sklyanin}
and the eigen-spectra is given by
\be
T(u)&=& (\Lambda_1(u)+\Lambda_2(u))/g(u) \label{T-ei}  \\
\Lambda_1(u)&=&\omega_1(u) \phi(u+\lam)Q(u-\lam)/Q(u) \\
 \Lambda_2(u)&=& \omega_2(u) \phi(u)Q(u+\lam)/Q(u)  \\
 g(u)&=&\sin(2u+\lambda)
\ee
where
\be
Q(u)&=&\disp\prod_{m=1}^{M} \{ \sin(u-v_m)
 \sin(u+v_m+\lam)\} \label{sixQ}\\
\phi(u)&=& \sin^{2N}(u)   \label{sixphi12}\\
\omega_2(u)&=&\omega_1(-u-\lam)=\sin(\xi_++u+\lam)
  \sin(\xi_--u-\lam)\sin(2u) \;.  \label{sixomega12}
\ee
These $v_1,v_2,\cdots,v_M$ satisfy the Bethe ansatz equations
\be
T(v_j)=0\;. \label{BAE}
\ee
It is easy to see that
\be
\Lambda_1(u+\lam)\Lambda_2(u)=f^1(u)  \label{con}\;.
\ee
Thus ansatz (\ref{T-ei}) solves the functional relations (\ref{Func-T}).
If we use the same ansatz (\ref{T-ei}) with all trig-functions $\sin$
replaced with elliptic function $h$,
\be
Q(u)&=&\disp\prod_{m=1}^{M} \{h(u-v_m)h(u+v_m+\lam)\} \label{eightQ}\\
\phi(u)&=& h^{2N}(u)   \label{eightphi12}\\
\omega_2(u)&=&\omega_1(-u-\lam)=h(\xi_++u+\lam)
  h(\xi_--u-\lam)\no \\
 && \h\h \times H(2u)\Theta(2u+2\lam) \;,
 \label{eightomega12} \\
 g(u)&=& h(2u+\lam)\Theta(0)
\ee
 then (\ref{con}) is still correct for the
transfer matrix $T(u)=T^{(1,1)}(u)$ of the eight-vertex model
with  diagonal $K$ matrices, which are given by setting $\mu_\pm=0$.
Thus the ansatz solves the functional
relations of the eight-vertex model.
Moreover it is reduces to the solution of the six-vertex model
as the elliptic nome $p\to 0$. These facts, following Reshetikhin
\cite{Reshetikhin:83}, support that
the above ansatz could indeed be the
eigenvalues of the eight-vertex transfer matrix. Therefore
the thermodynamics of the models could
be determined by solving the Bethe ansatz solutions (\ref{T-ei})
and (\ref{BAE}).

\noindent{\bf 5.3. RSOS model with open boundaries:}
In \cite{PeBe:95} the functional relations for the restricted
SOS model or ABF model with open boundaries have been constructed.
Here we have shown the functional relations of the unrestricted SOS
model and the correspondence between vertex $K$- and face $K$-matrices.
But we have not given the explicit form
of the face $K$ matrices by this
correspondence. This however is a very interesting problem. We may
obtain the face $K$ matrices of the ABF model by restricting the SOS
model. Also whether or not the face $K$ matrices following from this
correspondence are the same as the face $K$ matrices presented in
\cite{PeBe:95} is an open question and is presently
under investigation. It is obvious that the same idea can be applied to
higher rank JMO IRF models and Belavin's $Z_n$ vertex models.

The definition of the SOS transfer matrix is not unique. The definition
given in (\ref{V}) is different from the
one presented in \cite{PeBe:95}
because it seems that we cannot get one from
another one by using simple symmetries
like crossing symmetry of the face
weights. However, the definition given in (\ref{V}) follows directly
from the original formulation given by Sklyanin \cite{Sklyanin} and is
good to study a square lattice rotated by $45^0$ (see \cite{YuBa:94}),
which is the natural geometry for an integrable loop model with open
boundaries in statistical mechanics.
This further study will be helpful to generalize the
study of Baxter's eight-vertex model with periodic boundaries in
\cite{Baxter:73} to the model with open boundaries and  to
well understand the critical behaviour of the face models
\cite{Cardy:89,SaBa:89}.

\section*{Acknowledgements} This research has been supported by the
Australian Research Council. The author  thanks
Bo-Yu Hou and Paul A. Pearce for discussions and is also grateful to
Murray Batchelor for a critical reading of the manuscript.

\bigskip\bigskip

{\bigskip\rm\bf\noindent  Appendix~A:
   Transfer matrices and their graphs}
\setcounter{equation}{0}
\renewcommand{\theequation}{A.\arabic{equation}}\newline
Here we prove the commuting relation (\ref{TT}) for the fused
models graphically following the Sklyanin's arguments in the study of
the unfused six-vertex model \cite{Sklyanin}.
For the sake of clarity let
us represent the $R$- and $K$-matrices by the graphs,
\be
\setlength{\unitlength}{0.0115in}%
\begin{picture}(120,60)(60,345)
\put(120,375){\scr$1$\vector( 1, 0){ 60}}
\put(150,345){\scr$2$\vector( 0, 1){ 60}}
\put(141,363){\scr$u$}
\put( 50,372){$R^{12}(u)$ $=$}
\end{picture}
\setlength{\unitlength}{0.0115in}%
\begin{picture}(201,60)(30,714)
\put(108,714){\line( 1, 1){ 30}}
\put(138,744){\vector(-1, 1){ 15}}
\put(258,774){\line(-1,-1){ 30}}
\put(228,744){\vector( 1,-1){ 15}}
\put(123,759){\line(-1, 1){ 15}}
\put(243,729){\line( 1,-1){ 15}}
\put( 36,735){$;$}
\put( 57,740){\scr$K_-(u)\;=\hs{0.65}u$}
\put(150,735){$;$}
\put(174,740){\scr$\stackrel{t}{K}_+(u)\;=\hs{0.4}u$}
\end{picture}
\ee
Therefore we can represent the Yang-Baxter
equation (\ref{YBE}) graphically by
\be
\begin{picture}(183,63)(90,0)
\put(15,-3){
\setlength{\unitlength}{0.0105in}%
\begin{picture}(183,63)(69,717)
\put( 81,765){\vector(-1, 0){0}}
\put(147,765){\vector(1, 0){0}}
\put(135,774){\vector( 0,1){0}}
\put( 75,765){\line( 1, 0){ 15}}
\put( 90,765){\line( 1,-1){ 30}}
\put(120,735){\line( 1, 0){ 30}}
\put( 75,735){\line( 1, 0){ 12}}
\put( 87,735){\line( 6, 5){ 35.5}}
\put(122.5,765){\line( 1, 0){ 30}}
\put(135,780){\line( 0,-1){ 60}}
\put(243,765){\vector(1, 0){0}}
\put(177,765){\vector(-1, 0){0}}
\put(189,774){\vector( 0,1){0}}
\put(249,765){\line(-1, 0){ 15}}
\put(234,765){\line(-1,-1){ 30}}
\put(204,735){\line(-1, 0){ 30}}
\put(249,735){\line(-1, 0){ 12}}
\put(237,735){\line(-6, 5){ 35.5}}
\put(201.5,765){\line(-1, 0){ 30}}
\put(189,780){\line( 0,-1){ 60}}
\put(162,744){$=$}
\put( 96,753){\scr$u\!\!+\!\!v$}
\put(138,768){\scr$v$}
\put(128,737){\scr$u$}
\put(192,738){\scr$v$}
\put(180,768){\scr$u$}
\put(210,753){\scr$u\!\!+\!\!v$}
\put( 69,762){\small$1$}
\put( 69,732){\small$3$}
\put(190,717){\small$2$}
\put(129,717){\small$2$}
\put(252,762){\small$3$}
\put(252,732){\small$1$}
\end{picture}}
\put(187,20){or}
\put(215,-3){
\setlength{\unitlength}{0.0105in}%
\begin{picture}(183,63)(69,717)
\put(145,765){\vector(-1, 0){0}}
\put(145,735){\vector(-1, 0){0}}
\put(135,774){\vector( 0,1){0}}
\put( 75,765){\line( 1, 0){ 15}}
\put( 90,765){\line( 1,-1){ 30}}
\put(120,735){\line( 1, 0){ 30}}
\put( 75,735){\line( 1, 0){ 12}}
\put( 87,735){\line( 6, 5){ 35.5}}
\put(122.5,765){\line( 1, 0){ 30}}
\put(135,780){\line( 0,-1){ 60}}
\put(241,765){\vector(-1, 0){0}}
\put(241,735){\vector(-1, 0){0}}
\put(189,772){\vector( 0,1){0}}
\put(249,765){\line(-1, 0){ 15}}
\put(234,765){\line(-1,-1){ 30}}
\put(204,735){\line(-1, 0){ 30}}
\put(249,735){\line(-1, 0){ 12}}
\put(237,735){\line(-6, 5){ 35.5}}
\put(201.5,765){\line(-1, 0){ 30}}
\put(189,780){\line( 0,-1){ 60}}
\put(162,744){$=$}
\put( 96,753){\scr$u\!\!-\!\!v$}
\put(138,768){\scr$u$}
\put(128,737){\scr$v$}
\put(192,738){\scr$u$}
\put(180,768){\scr$v$}
\put(210,753){\scr$u\!\!-\!\!v$}
\put( 69,762){\small$2$}
\put( 69,732){\small$1$}
\put(190,717){\small$3$}
\put(129,717){\small$3$}
\put(252,762){\small$1$}
\put(252,732){\small$2$}
\end{picture}}\end{picture}
\label{YB}
\ee
the reflection equations (\ref{ref-}) and (\ref{ref+}) by
\be
\begin{picture}(183,93)(90,-30)
\put(25,90){
\setlength{\unitlength}{0.0085in}%
\begin{picture}(183,63)(69,717)
\put(120,675){\line( 2,-3){ 60}}
\put(180,585){\line(-2,-3){ 30}}
\put(117,648){\line( 2,-1){ 63.600}}
\put(180,615){\line(-2,-1){ 63.600}}
\put(180,675){\line( 0,-1){135}}
\put(213,633){\line( 2,-1){ 63.600}}
\put(276,600){\line(-2,-1){ 63.600}}
\put(216,540){\line( 2, 3){ 60}}
\put(276,630){\line(-2, 3){ 30}}
\put(276,675){\line( 0,-1){135}}
\put(198,606){$=$}
\put(168,506){for (\ref{ref-})}
\put(225.7,555){\vector( 3, 4){  0}}
\put(234,579){\vector( 3, 2){0}}
\put(144,598){\vector( 3, 2){0}}
\put(165,561){\vector( 1, 1){0}}
\put(111,579){\small$1$}
\put(144,540){\small$2$}
\put(224,540){\small$2$}
\put(207,570){\small$1$}
\put(153,594){\scr$u\!\!+\!\!v$}
\put(252,612){\scr$u\!\!+\!\!v$}
\put(135,636){\scr$u\!\!-\!\!v$}
\put(231,570){\scr$u\!\!-\!\!v$}
\put(172,565){\scr$v$}
\put(171,621){\scr$u$}
\put(268,588){\scr$u$}
\put(268,641){\scr$v$}
\put(347,603){and}
\end{picture}}
\put(95,90){
\setlength{\unitlength}{0.0085in}%
\begin{picture}(183,63)(69,717)
\put(380.2,555){\vector(3,-4){0}}
\put(372,579){\vector(3,-2){0}}
\put(471,591.4){\vector(3,-2){0}}
\put(441,562){\vector(1,-1){0}}
\put(330,675){\line( 0,-1){135}}
\put(426,675){\line( 0,-1){135}}
\put(486,675){\line(-2,-3){ 60}}
\put(426,585){\line( 2,-3){ 30}}
\put(489,648){\line(-2,-1){ 63.600}}
\put(426,615){\line( 2,-1){ 63.600}}
\put(393,633){\line(-2,-1){ 63.600}}
\put(330,600){\line( 2,-1){ 63.600}}
\put(390,540){\line(-2, 3){ 60}}
\put(330,630){\line( 2, 3){ 30}}
\put(495,579){\small$1$}
\put(462,540){\small$2$}
\put(386,542){\small$2$}
\put(399,570){\small$1$}
\put(400,606){$=$}
\put(370,506){for (\ref{ref+})}
\put(342,612){\scr$-\!\!u\!\!-\!\!v\!\!-\!\!2\!\lam$}
\put(435,597){\scr$-\!\!u\!\!-\!\!v\!\!-\!\!2\!\lam$}
\put(465,639){\scr$v\!\!-\!\!u$}
\put(351,588){\scr$v\!\!-\!\!u$}
\put(429,567){\scr$v$}
\put(429,621){\scr$u$}
\put(330,636){\scr$v$}
\put(333,588){\scr$u$}
\end{picture}}
\end{picture}
\label{RYB}
\ee
and the unitary and crossing unitary relations
by
\be
\setlength{\unitlength}{0.0115in}%
\begin{picture}(144,36)(137,372)
\put(126,405){\line( 1, 0){ 24}}
\put(150,405){\line( 1,-1){ 30}}
\put(180,375){\line( 1, 0){ 30}}
\put(210,375){\line( 1, 1){ 30}}
\put(240,405){\line( 1, 0){ 21}}
\put(261,405){\line(-1, 0){  3}}
\put(126,375){\line( 1, 0){ 24}}
\put(150,375){\line( 1, 1){ 30}}
\put(180,405){\line( 1, 0){ 27}}
\put(207,405){\line( 6,-5){ 35.5}}
\put(242.7,375){\line( 1, 0){ 21}}
\put(192,375){\vector( 1, 0){ 0}}
\put(189,405){\vector( 1, 0){ 0}}
\put(156,387){\scr$u$}
\put(210,387){\scr$-\!u$}
\put(117,372){\small$2$}
\put(117,399){\small$1$} \put(300,387){$=$}
\put(330,387){$\rho(u)$}
\end{picture}\label{I}\ee
\be
\setlength{\unitlength}{0.0115in}%
\begin{picture}(144,36)(137,372)
\put(126,405){\line( 1, 0){ 24}}
\put(150,405){\line( 1,-1){ 30}}
\put(180,375){\line( 1, 0){ 30}}
\put(210,375){\line( 1, 1){ 30}}
\put(240,405){\line( 1, 0){ 21}}
\put(261,405){\line(-1, 0){  3}}
\put(126,375){\line( 1, 0){ 24}}
\put(150,375){\line( 1, 1){ 30}}
\put(180,405){\line( 1, 0){ 27}}
\put(207,405){\line( 6,-5){ 35.5}}
\put(242.7,375){\line( 1, 0){ 21}}
\put(192,375){\vector(1, 0){ 0}}
\put(191,405){\vector(-1, 0){ 0}}
\put(134,387){\scr$-\!u\!\!-\!\!2\!\lam$}
\put(214,387){\scr$u$}
\put(117,372){\small$2$}
\put(117,399){\small$1$} \put(300,387){$=$}
\put(330,387){$\tilde{\rho}_{1,1}(u)\;.$}
\end{picture}\label{cross-I}\ee

Using the graphs the fused $R$ matrix $R_{(p,q)}(u)$ (\ref{Rpq})
is given by the elements
\be
\setlength{\unitlength}{0.0115in}%
\begin{picture}(75,84)(123,673)
\thicklines
\put(105,735){\vector( 1, 0){ 60}}
\put(135,705){\vector( 0, 1){ 60}}
\put(109,740){\small$p$}
\put(137,709){\small$q$}
\put( 93,729){$i$}
\put(168,729){$j\hs{0.3}=$}
\put(135,768){$l$}
\put(127,726){$u$}
\put(132,693){$k$}
\end{picture}
\setlength{\unitlength}{0.0115in}%
\begin{picture}(210,132)(25,660)
\multiput(150,720)(9.60000,0.00000){6}{
  \makebox(0.4444,0.6667){\tenrm .}}
\multiput(171,702)(0.00000,9.00000){5}{
  \makebox(0.4444,0.6667){\tenrm .}}
\put( 75,690){\vector( 1, 0){210}}
\put( 75,750){\vector( 1, 0){210}}
\put( 75,765){\vector( 1, 0){210}}
\put( 90,660){\vector( 0, 1){120}}
\put(105,660){\vector( 0, 1){120}}
\put(255,660){\vector( 0, 1){120}}
\put(270,660){\vector( 0, 1){120}}
\put( 75,675){\vector( 1, 0){210}}
\put(249,683){\scr$u_{22}$}
\put(265,768){\scr$u_{p1}$}
\put(268,683){\scr$u_{21}$}
\put(268,666){\scr$u_{11}$}
\put(249,666){\scr$u_{12}$}
\put(85,768){\scr$u_{pq}$}
\put(85,666){\scr$u_{1q}$}
\put( 66,762){\scr$i_p$}
\put( 66,687){\scr$i_2$}
\put( 66,670){\scr$i_1$}
\put(288,670){\scr$j_1$}
\put(288,687){\scr$j_2$}
\put(291,762){\scr$j_p$}
\put( 87,650){\scr$k_1$}
\put(267,650){\scr$k_q$}
\put(102,650){\scr$k_2$}
\put(105,783){\scr$l_2$}
\put(270,783){\scr$l_q$}
\put( 87,783){\scr$l_1$}
\put(40,717){\scr$\disp{\sum_{\bf j,l}}$}
\end{picture}
\ee
and the fused $K$  matrices $K_-^{(q)}(u)$  (\ref{fK-})
and $\stackrel{t}{K}_+^{(q)}(u)$ (\ref{fK+}) are given, respectively,
 by the elements
\be
\setlength{\unitlength}{0.0115in}%
\begin{picture}(57,81)(154,553)
\thicklines
\put(120,645){\line( 1, 1){ 30}}
\put(150,675){\vector(-1, 1){ 30}}
\put(153,672){$u$}
\put(114,705){$j$}
\put(114,633){$i$}
\put(150,700){\thinlines\line( 0,-1){60}}
\put(171,669){$=\;\rho(u|q)\;\;$\scr$\disp{\sum_{\bf j}}$}
\end{picture}
\setlength{\unitlength}{0.0115in}%
\begin{picture}(123,237)(78,543)
\put(269,599){\line(-1,-1){ 30}}
\multiput(255,669)(0.00000,-2.00000){17}{\line( 0,-1){  1.000}}
\put(240,750){\line( 1,-1){ 30}}
\put(255,765){\line( 1,-1){ 15}}
\put(240,720){\line(-1,-1){ 15}}
\put(180,690){\line( 1,-1){ 90}}
\put(165,675){\line(1, -1){ 105}}
\put(270,570){\line(-1,-1){ 15}}
\put(270,750){\line(-1,-1){ 105}}
\put(180,630){\line( 1, 1){ 90}}
\put(270,780){\line( 0,-1){237}}
\put(255,765){\vector(-1, 1){0}}
\put(165,675){\vector(-1, 1){0}}
\put(180,690){\vector(-1, 1){0}}
\put(240,750){\vector(-1, 1){0}}
\put(276,564){\scr$u$}
\put(273,744){\scr$u\!\!+\!\!q\!\lam\!\!-\!\!\lam$}
\put(276,597){\scr$u\!\!+\!\!\lam$}
\put(156,657){\scr$2u\!\!+\!\!q\!\lam\!\!-\!\!\lam$}
\put(237,579){\scr$2u\!\!+\!\!\lam$}
\put(180,636){\scr$2u\!\!+\!\!q\!\lam\!\!-\!\!2\!\lam$}
\put(189,678){\scr$2u\!\!+\!\!q\!\lam$}
\put(231,726){\scr$2u\!\!+\!\!2\!q\!\lam\!\!-\!\!3\!\lam$}
\put(174,615){\scr$i_2$}
\put(246,543){\scr$i_q$}
\put(156,633){\scr$i_1$}
\put(153,678){\scr$j_q$}
\put(171,696){\scr$j_{q-1}$}
\put(252,768){\scr$j_1$}
\end{picture}
\ee
and
\be
\setlength{\unitlength}{0.0115in}%
\begin{picture}(63,75)(218,609)
\thicklines
\put(240,765){\line(-1,-1){ 30}}
\put(210,735){\vector( 1,-1){ 30}}
\put(247,699){$j$}
\put(246,765){$i$}
\put(210,763){\thinlines\line( 0,-1){60}}
\put(219,732){$u$}
\put(273,732){$=\;\;$\scr$\disp{\sum_{\bf j}}$}
\end{picture}
\setlength{\unitlength}{0.0115in}%
\begin{picture}(144,240)(80,579)
\put(243,729){\line(-1,-1){ 45}}
\put(153,579){\line( 0, 1){237}}
\multiput(255,687)(0.42857,-0.42857){8}{
 \makebox(0.4444,0.6667){\sevrm .}}
\put(258,684){\vector( 1,-1){0}}
\multiput(240,672)(0.42857,-0.42857){8}{
 \makebox(0.4444,0.6667){\sevrm .}}
\put(243,669){\vector( 1,-1){0}}
\multiput(180,612)(0.42857,-0.42857){8}{
 \makebox(0.4444,0.6667){\sevrm .}}
\put(183,609){\vector( 1,-1){0}}
\multiput(171,681)(0.00000,2.00000){17}{\line( 0, 1){  1.000}}
\put(168,804){\line(-1,-1){ 15}}
\put(153,789){\line( 1,-1){102}}
\put(183,789){\line(-1,-1){30}}
\put(153,759){\line( 1,-1){ 87}}
\put(153,639){\line( 1, 1){ 90}}
\put(153,609){\line( 1, 1){105}}
\put(153,609){\vector( 1,-1){ 15}}
\put(153,639){\line( 1,-1){ 27}}
\put(127,636){\scr$u\!+\!\lam$}
\put(141,603){\scr$u$}
\put(117,783){\scr$u\!\!+\!\!q\!\lam\!\!-\!\!\lam$}
\put(172,807){\scr$i_1$}
\put(168,585){\scr$j_q$}
\put(243,660){\scr$j_2$}
\put(240,738){\scr$i_{q-1}$}
\put(261,720){\scr$i_q$}
\put(261,673){\scr$j_1$}
\put(159,618){\scr$-\!\!2u\!\!-\!\!3\lam$}
\put(210,717){\scr$-\!\!2u\!\!-\!\!q\!\lam\!\!-\!\!2\!\lam$}
\put(237,699){\scr$-\!\!2u\!\!-\!\!q\!\lam\!\!-\!\!\lam$}
\put(213,675){\scr$-\!\!2u\!\!-\!\!q\!\lam$}
\put(162,765){\scr$-\!\!2u\!\!-\!\!2q\!\lam\!\!+\!\!\lam$}
\end{picture}
\ee
where $u_{r,s}=u+r\lam-s\lam$ and
sum over $\bf j,l$ is over all possible spins $j_1,j_2,\cdots,j_p$ and
$l_1,l_2,\cdots,l_q$ satisfying
$$j=\sum_{s=1}^pj_s \and l=\sum_{s=1}^ql_s\;.$$
The spins $i$ and $k$ are
$$i=\sum_{s=1}^pi_s \and k=\sum_{s=1}^qk_s\;.$$
The indices $i,j=-\phalf,-\phalf+1,\cdots,\phalf$ and
$k,l=-\qhalf,-\qhalf+1,\cdots,\qhalf$.

The fused Yang-Baxter equation (\ref{YBE}) and reflection
equations (\ref{fref-})-(\ref{fref+}) are expressed by graphs
\be
\begin{picture}(183,63)(90,0)
\put(15,-3){
\setlength{\unitlength}{0.0105in}%
\begin{picture}(183,63)(69,717)
\thicklines
\put( 81,765){\vector(-1, 0){0}}
\put(147,765){\vector(1, 0){0}}
\put(135,774){\vector( 0,1){0}}
\put( 75,765){\line( 1, 0){ 15}}
\put( 90,765){\line( 1,-1){ 30}}
\put(120,735){\line( 1, 0){ 30}}
\put( 75,735){\line( 1, 0){ 12}}
\put( 87,735){\line( 6, 5){ 35.5}}
\put(122.5,765){\line( 1, 0){ 30}}
\put(135,780){\line( 0,-1){ 60}}
\put(243,765){\vector(1, 0){0}}
\put(177,765){\vector(-1, 0){0}}
\put(189,774){\vector( 0,1){0}}
\put(249,765){\line(-1, 0){ 15}}
\put(234,765){\line(-1,-1){ 30}}
\put(204,735){\line(-1, 0){ 30}}
\put(249,735){\line(-1, 0){ 12}}
\put(237,735){\line(-6, 5){ 35.5}}
\put(201.5,765){\line(-1, 0){ 30}}
\put(189,780){\line( 0,-1){ 60}}
\put(162,744){$=$}
\put( 96,753){\scr$u\!\!+\!\!v$}
\put(138,768){\scr$v$}
\put(128,737){\scr$u$}
\put(192,738){\scr$v$}
\put(180,768){\scr$u$}
\put(210,753){\scr$u\!\!+\!\!v$}
\put( 69,762){\small$p$}
\put( 69,732){\small$b$}
\put(190,717){\small$q$}
\put(129,717){\small$q$}
\put(252,762){\small$b$}
\put(252,732){\small$p$}
\end{picture}}
\put(187,20){or}
\put(215,-3){
\setlength{\unitlength}{0.0105in}%
\begin{picture}(183,63)(69,717)
\thicklines
\put(145,765){\vector(-1, 0){0}}
\put(145,735){\vector(-1, 0){0}}
\put(135,774){\vector( 0,1){0}}
\put( 75,765){\line( 1, 0){ 15}}
\put( 90,765){\line( 1,-1){ 30}}
\put(120,735){\line( 1, 0){ 30}}
\put( 75,735){\line( 1, 0){ 12}}
\put( 87,735){\line( 6, 5){ 35.5}}
\put(122.5,765){\line( 1, 0){ 30}}
\put(135,780){\line( 0,-1){ 60}}
\put(241,765){\vector(-1, 0){0}}
\put(241,735){\vector(-1, 0){0}}
\put(189,774){\vector( 0,1){0}}
\put(249,765){\line(-1, 0){ 15}}
\put(234,765){\line(-1,-1){ 30}}
\put(204,735){\line(-1, 0){ 30}}
\put(249,735){\line(-1, 0){ 12}}
\put(237,735){\line(-6, 5){ 35.5}}
\put(201.5,765){\line(-1, 0){ 30}}
\put(189,780){\line( 0,-1){ 60}}
\put(162,744){$=$}
\put( 96,753){\scr$u\!\!-\!\!v$}
\put(138,768){\scr$u$}
\put(128,737){\scr$v$}
\put(192,738){\scr$u$}
\put(180,768){\scr$v$}
\put(210,753){\scr$u\!\!-\!\!v$}
\put( 69,762){\small$q$}
\put( 69,732){\small$p$}
\put(190,717){\small$b$}
\put(129,717){\small$b$}
\put(252,762){\small$p$}
\put(252,732){\small$q$}
\end{picture}}\end{picture}
\label{FYB}
\ee
\be
\begin{picture}(183,93)(90,-30)
\thicklines
\put(25,90){
\setlength{\unitlength}{0.0085in}%
\begin{picture}(183,63)(69,717)
\put(120,675){\line( 2,-3){ 60}}
\put(180,585){\line(-2,-3){ 30}}
\put(117,648){\thinlines\line( 2,-1){ 63.600}}
\put(117,649){\thinlines\line( 2,-1){ 63.600}}
\put(180,615){\thinlines\line(-2,-1){ 63.600}}
\put(180,614){\thinlines\line(-2,-1){ 63.600}}
\put(180,675){\thinlines\line( 0,-1){135}}
\put(213,633){\thinlines\line( 2,-1){ 63.600}}
\put(213,634){\thinlines\line( 2,-1){ 63.600}}
\put(276,600){\thinlines\line(-2,-1){ 63.600}}
\put(276,599){\thinlines\line(-2,-1){ 63.600}}
\put(216,540){\line( 2, 3){ 60}}
\put(276,630){\line(-2, 3){ 30}}
\put(276,675){\thinlines\line( 0,-1){135}}
\put(198,606){$=$}
\put(168,506){for (\ref{fref-})}
\put(225.7,555){\vector( 3, 4){  0}}
\put(234,579){\vector( 3, 2){0}}
\put(144,598){\vector( 3, 2){0}}
\put(165,561){\vector( 1, 1){0}}
\put(111,579){\small$p$}
\put(144,540){\small$q$}
\put(224,540){\small$q$}
\put(207,570){\small$p$}
\put(153,594){\scr$u\!\!+\!\!v\!\!-\!\!\lam\!\!+\!\!p\!\lam$}
\put(252,612){\scr$u\!\!+\!\!v\!\!-\!\!\lam\!\!+\!\!p\!\lam$}
\put(135,636){\scr$u\!\!-\!\!v$}
\put(231,570){\scr$u\!\!-\!\!v$}
\put(172,567){\scr$v$}
\put(172,621){\scr$u$}
\put(268,588){\scr$u$}
\put(268,639){\scr$v$}
\put(347,603){and}
\end{picture}}
\put(95,90){
\setlength{\unitlength}{0.0085in}%
\begin{picture}(183,63)(69,717)
\put(380.2,555){\vector(3,-4){0}}
\put(372,579){\vector(3,-2){0}}
\put(477,589.4){\vector(3,-2){0}}
\put(445,557){\vector(1,-1){0}}
\put(330,675){\thinlines\line( 0,-1){135}}
\put(426,675){\thinlines\line( 0,-1){135}}
\put(486,675){\line(-2,-3){ 60}}
\put(426,585){\line( 2,-3){ 30}}
\put(489,648){\thinlines\line(-2,-1){ 63.600}}
\put(489,647){\thinlines\line(-2,-1){ 63.600}}
\put(426,615){\thinlines\line( 2,-1){ 63.600}}
\put(426,616){\thinlines\line( 2,-1){ 63.600}}
\put(393,633){\thinlines\line(-2,-1){ 63.600}}
\put(393,632){\thinlines\line(-2,-1){ 63.600}}
\put(330,600){\thinlines\line( 2,-1){ 63.600}}
\put(330,601){\thinlines\line( 2,-1){ 63.600}}
\put(390,540){\line(-2, 3){ 60}}
\put(330,630){\line( 2, 3){ 30}}
\put(495,579){\small$p$}
\put(462,540){\small$q$}
\put(386,542){\small$q$}
\put(399,570){\small$p$}
\put(400,600){$=$}
\put(370,506){for (\ref{fref+})}
\put(342,612){\scr$-\!\!u\!\!-\!\!v\!\!-\!\!\lam\!\!-\!\!p\!\lam$}
\put(435,597){\scr$-\!\!u\!\!-\!\!v\!\!-\!\!\lam\!\!-\!\!p\!\lam$}
\put(465,639){\scr$v\!\!-\!\!u$}
\put(351,588){\scr$v\!\!-\!\!u$}
\put(428,569){\scr$v$}
\put(429,621){\scr$u$}
\put(328.5,636){\scr$v$}
\put(333,588){\scr$u$}
\end{picture}}
\end{picture}
\label{fRYB}
\ee
The unitary (\ref{funity}) and crossing unitary
conditions (\ref{fcross-unity}) for the fused $R$
matrix become
\be
\setlength{\unitlength}{0.0115in}%
\begin{picture}(144,36)(137,372)\thicklines
\put(126,405){\line( 1, 0){ 24}}
\put(150,405){\line( 1,-1){ 30}}
\put(180,375){\line( 1, 0){ 30}}
\put(210,375){\line( 1, 1){ 30}}
\put(240,405){\line( 1, 0){ 21}}
\put(261,405){\line(-1, 0){  3}}
\put(126,375){\line( 1, 0){ 24}}
\put(150,375){\line( 1, 1){ 30}}
\put(180,405){\line( 1, 0){ 27}}
\put(207,405){\line( 6,-5){ 35.5}}
\put(242.7,375){\line( 1, 0){ 21}}
\put(192,375){\vector( 1, 0){ 0}}
\put(189,405){\vector( 1, 0){ 0}}
\put(156,387){\scr$u$}
\put(210,387){\scr$-\!u$}
\put(117,372){\small$q$}
\put(117,399){\small$p$} \put(300,387){$=$}
\put(330,387){$\rho_{p,q}(u)Y^+_pY^+_q$}
\end{picture}\label{fI}\ee
\be
\setlength{\unitlength}{0.0115in}%
\begin{picture}(144,36)(137,372)\thicklines
\put(126,405){\line( 1, 0){ 24}}
\put(150,405){\line( 1,-1){ 30}}
\put(180,375){\line( 1, 0){ 30}}
\put(210,375){\line( 1, 1){ 30}}
\put(240,405){\line( 1, 0){ 21}}
\put(261,405){\line(-1, 0){  3}}
\put(126,375){\line( 1, 0){ 24}}
\put(150,375){\line( 1, 1){ 30}}
\put(180,405){\line( 1, 0){ 27}}
\put(207,405){\line( 6,-5){ 35.5}}
\put(242.7,375){\line( 1, 0){ 21}}
\put(192,375){\vector(1, 0){ 0}}
\put(191,405){\vector(-1, 0){ 0}}
\put(134,387){\scr$-\!u\!\!-\!\!2\!\lam$}
\put(214,387){\scr$u$}
\put(117,372){\small$q$}
\put(117,399){\small$p$} \put(300,387){$=$}
\put(330,387){$\tilde{\rho}_{q,p}(u)Y^+_pY^+_q\;$}
\end{picture}\label{fcross-I}\ee

To prove (\ref{TT})  consider $\T^{(p,q)}(u)\T^{(p,b)}(v)$
\be\setlength{\unitlength}{0.0125in}%
\begin{picture}(186,120)(48,675)\thicklines
\put(105,735){\vector( 0, 1){ 60}}
\put(120,735){\vector( 0, 1){ 60}}
\put(195,735){\vector( 0, 1){ 60}}
\put(180,735){\vector( 0, 1){ 60}}
\multiput(138,765)(8.00000,0.00000){4}{
 \makebox(0.4444,0.6667){\tenrm .}}
\put(150,690){\vector( 1, 0){  6}}
\put(105,675){\vector( 0, 1){ 60}}
\put(120,675){\vector( 0, 1){ 60}}
\put(195,675){\vector( 0, 1){ 60}}
\put(180,675){\vector( 0, 1){ 60}}
\multiput(138,705)(8.00000,0.00000){4}{
 \makebox(0.4444,0.6667){\tenrm .}}
\put(129,720){\line(-1, 0){ 39}}
\put( 90,720){\line(-1,-1){ 15}}
\put( 75,705){\line( 1,-1){ 15}}
\put( 90,690){\line( 1, 0){120}}
\put(210,690){\line( 1, 1){ 15}}
\put(225,705){\line(-1, 1){ 15}}
\put(210,720){\vector(-1, 0){ 84}}
\put( 75,792){\thinlines\line( 0,-1){114}}
\put(225,795){\thinlines\line( 0,-1){114}}
\put(129,780){\line(-1, 0){ 39}}
\put( 90,780){\line(-1,-1){ 15}}
\put( 75,765){\line( 1,-1){ 15}}
\put( 90,750){\line( 1, 0){120}}
\put(210,750){\line( 1, 1){ 15}}
\put(225,765){\line(-1, 1){ 15}}
\put(210,780){\vector(-1, 0){ 84}}
\put(159,750){\vector( 1, 0){  6}}
\put(234,762){\scr$K_-^{(b)}(v)$}
\put(42,762){\scr$K_+^{(b)}(v)$}
\put(195,773){\scr$v\!\!+\!\!b\!\lam\!\!-\!\!\lam$}
\put(234,702){\scr$K_-^{(q)}(u)$}
\put(42,702){\scr$K_+^{(q)}(u)$}
\put( 99,681){\scr$u$}
\put(198,711){\scr$u\!\!+\!\!q\!\lam\!\!-\!\!\lam$}
\put( 96,747){\small$b$}
\put( 96,717){\small$a$}
\put(168,743){\scr$v$}
\end{picture}\label{T(u)T(v)}\ee
Inserting the operators $Y^+_q$ and $Y^+_b$ respectively into
the positions $a$ and $b$ and using the crossing unitary
condition (\ref{fcross-I}) and then
using the Yang-Baxter equation (\ref{FYB}) to push the vertex with
spectra parameter $u+v-\lam+b\lam$ through to the most right, we are
able to obtain
\be\setlength{\unitlength}{0.0125in}%
\begin{picture}(192,105)(-10,535)\thicklines
\put( 75,630){\thinlines\line( 0,-1){90}}
\put(225,630){\thinlines\line( 0,-1){90}}
\put( 75,570){\line( 1, 1){ 30}}
\put(105,600){\line( 1, 0){ 90}}
\put(195,600){\line( 1,-1){ 30}}
\put(225,570){\line(-1,-1){ 15}}
\put(210,555){\line(-1, 0){120}}
\put( 90,555){\line(-1, 1){ 15}}
\put(111,540){\vector( 0, 1){90}}
\put(126,540){\vector( 0, 1){90}}
\put(186,540){\vector( 0, 1){90}}
\put(171,540){\vector( 0, 1){90}}
\multiput(138,585)(8.00000,0.00000){4}{
 \makebox(0.4444,0.6667){\tenrm .}}
\put(144,555){\vector( 1, 0){  6}}
\put(144,570){\vector( 1, 0){  6}}
\put(144,615){\vector(-1, 0){  3}}
\put(147,600){\vector(-1, 0){  3}}
\put( 75,600){\line( 1,-1){ 30}}
\put(105,570){\line( 1, 0){ 90}}
\put(195,570){\line( 1, 1){ 30}}
\put(225,600){\line(-1, 1){ 15}}
\put(210,615){\line(-1, 0){120}}
\put( 90,615){\line(-1,-1){ 15}}
\put(42,597){\scr$K_+^{(b)}(v)$}
\put(42,552){\scr$K_+^{(q)}(u)$}
\put(-58,578){\small$\disp{1\over  \tilde{\rho}_{q,b}
  (u\!\!+\!\!v\!\!-\!\!\lam\!\!+\!\!b\!\lam)}\;\times$}
\put(231,552){\scr$K_-^{(q)}(u)$}
\put(231,597){\scr$K_-^{(b)}(v)$}
\put(105,546){\scr$u$}
\put(186,608){\scr$v\!\!+\!\!b\!\lam\!\!-\!\!\lam$}
\put(125,594){\scr$u\!\!+\!\!q\!\lam\!\!-\!\!\lam$}
\put(105,563){\scr$v$}
\put( 76,588){\scr$-\!v\!\!-\!\!u\!\!-\!\!\lam\!\!-\!\!b\!\lam$}
\put(198,573){\scr$u\!\!+\!\!v\!\!-\!\!\lam\!\!+\!\!b\!\lam$}
\put(102,600){\small$a$}
\put( 96,615){\small$c$}
\end{picture}\label{C8}
\ee
Then inserting the operators $Y^+_q$ and $Y^+_b$ again into
the positions $a$ and $c$ respectively and
employing  the unitary condition
(\ref{fI})  and then using the Yang-Baxter
equation (\ref{FYB}) to push the vertex with spectra parameter
$v-u$ through again to the most right, we have
\be
\setlength{\unitlength}{0.0085in}%
\begin{picture}(291,120)(-83,675)\thicklines
\put( 60,780){\thinlines\line( 0,-1){ 90}}
\put( 60,750){\line( 1, 1){ 30}}
\put( 90,780){\line( 1,-1){ 30}}
\put(120,780){\line(-1,-1){ 30}}
\put( 90,750){\line(-1,-1){ 30}}
\put( 60,720){\line( 1,-1){ 30}}
\put( 90,690){\line( 1, 0){ 81}}
\put( 60,750){\line( 1,-1){ 30}}
\put( 90,720){\line( 1, 0){ 27}}
\put(117,720){\line( 1, 0){ 51}}
\put(120,750){\line( 1, 0){ 48}}
\put(120,780){\line( 1, 0){ 48}}
\put(321,780){\thinlines\line( 0,-1){ 90}}
\put(321,750){\line(-1, 1){ 30}}
\put(291,780){\line(-1,-1){ 30}}
\put(261,780){\line( 1,-1){ 30}}
\put(291,750){\line( 1,-1){ 30}}
\put(321,720){\line(-1,-1){ 30}}
\put(291,690){\line(-1, 0){ 81}}
\put(321,750){\line(-1,-1){ 30}}
\put(291,720){\line(-1, 0){ 27}}
\put(264,720){\line(-1, 0){ 51}}
\put(261,750){\line(-1, 0){ 48}}
\put(261,780){\line(-1, 0){ 48}}
\put(165,780){\line( 1, 0){ 48}}
\put(165,750){\line( 1, 0){ 48}}
\put(165,720){\line( 1, 0){ 48}}
\put(168,690){\line( 1, 0){ 48}}
\put(135,675){\vector( 0, 1){120}}
\put(156,675){\vector( 0, 1){120}}
\put(249,675){\vector( 0, 1){120}}
\put(228,675){\vector( 0, 1){120}}
\put(192,750){\vector(-1, 0){  3}}
\put(195,780){\vector(-1, 0){  3}}
\put(186,690){\vector( 1, 0){  6}}
\multiput(177,735)(10.00000,0.00000){4}{
 \makebox(0.4444,0.6667){\tenrm .}}
\put(188,720){\vector( 1, 0){0}}
\put( 96,765){\scr$u\!-\!v$}
\put(267,765){\scr$v\!-\!u$}
\put(270,730){\scr$v\!\!+\!\!u\!\!-\!\!\lam\!\!+\!\!b\!\lam$}
\put(127,708){\scr$v$}
\put(157,741){\scr$v\!\!+\!\!b\!\lam\!\!-\!\!\lam$}
\put(157,771){\scr$u\!\!+\!\!q\!\lam\!\!-\!\!\lam$}
\put(127,681){\scr$u$}
\put(17,744){\scr$K_+^{(b)}(v)$}
\put(17,714){\scr$K_+^{(q)}(u)$}
\put(324,714){\scr$K_-^{(q)}(u)$}
\put(324,747){\scr$K_-^{(b)}(v)$}
\put( 63,738){\scr$-\!v\!\!-\!\!u\!\!-\!\!\lam\!\!-\!\!b\!\lam$}
\put(-200,730){\small$\disp{1\over
 \tilde{\rho}_{q,b}(u\!\!+\!\!v\!\!-\!\!\lam\!\!+\!\!b\!\lam)
 \rho_{b,q}(u\!\!-\!\!v)}\;\times$}
\end{picture}
\ee
By the reflection equations (\ref{fRYB}) we obtain
\be
\setlength{\unitlength}{0.0085in}%
\begin{picture}(294,120)(-80,525)\thicklines
\put( 60,540){\thinlines\line( 0, 1){ 90}}
\put( 60,570){\line( 1,-1){ 30}}
\put( 90,540){\line( 1, 1){ 30}}
\put(120,540){\line(-1, 1){ 30}}
\put( 90,570){\line(-1, 1){ 30}}
\put( 60,600){\line( 1, 1){ 30}}
\put( 90,630){\line( 1, 0){ 81}}
\put( 60,570){\line( 1, 1){ 30}}
\put( 90,600){\line( 1, 0){ 27}}
\put(117,600){\line( 1, 0){ 51}}
\put(120,570){\line( 1, 0){ 48}}
\put(120,540){\line( 1, 0){ 48}}
\put(321,540){\thinlines\line( 0, 1){ 90}}
\put(321,570){\line(-1,-1){ 30}}
\put(291,540){\line(-1, 1){ 30}}
\put(261,540){\line( 1, 1){ 30}}
\put(291,570){\line( 1, 1){ 30}}
\put(321,600){\line(-1, 1){ 30}}
\put(291,630){\line(-1, 0){ 81}}
\put(321,570){\line(-1, 1){ 30}}
\put(291,600){\line(-1, 0){ 27}}
\put(264,600){\line(-1, 0){ 51}}
\put(261,570){\line(-1, 0){ 48}}
\put(261,540){\line(-1, 0){ 48}}
\put(165,540){\line( 1, 0){ 48}}
\put(165,570){\line( 1, 0){ 48}}
\put(165,600){\line( 1, 0){ 48}}
\put(168,630){\line( 1, 0){ 48}}
\multiput(177,585)(10.00000,0.00000){4}{
 \makebox(0.4444,0.6667){\tenrm .}}
\put(192,600){\vector(-1, 0){  3}}
\put(186,570){\vector( 1, 0){  6}}
\put(192,630){\vector(-1, 0){  3}}
\put(183,540){\vector( 1, 0){  9}}
\put(135,525){\vector( 0, 1){120}}
\put(156,525){\vector( 0, 1){120}}
\put(228,525){\vector( 0, 1){120}}
\put(249,525){\vector( 0, 1){120}}
\put(17,597){\scr$K_+^{(q)}(u)$}
\put(324,564){\scr$K_-^{(b)}(v)$}
\put(17,567){\scr$K_+^{(b)}(v)$}
\put( 96,558){\scr$u\!-\!v$}
\put(267,558){\scr$v\!-\!u$}
\put(270,584){\scr$v\!\!+\!\!u\!\!-\!\!\lam\!\!+\!\!b\!\lam$}
\put(127,531){\scr$u$}
\put(157,621){\scr$u\!\!+\!\!q\!\lam\!\!-\!\!\lam$}
\put(127,561){\scr$v$}
\put(157,591){\scr$v\!\!+\!\!b\!\lam\!\!-\!\!\lam$}
\put(324,597){\scr$K_-^{(q)}(u)$}
\put( 60,588){\scr$-\!v\!\!-\!\!u\!\!-\!\lam\!\!-\!\!b\!\lam$}
\put(-200,580){\small$\disp{1\over
\tilde{\rho}_{q,b}(u\!\!+\!\!v\!\!-\!\!\lam\!\!+\!\!b\!\lam)
 \rho_{b,q}(u\!\!-\!\!v)}\;\times$}
\end{picture}
\ee
Using the Yang-Baxter equation to push the vertex with
spectral parameter $v-u$ back through  to the left and removing
the vertices with spectral parameters $v-u$ and $u-v$ by
the unitary condition (\ref{fI}), then pushing the vertex with
spectral parameter $v+u-\lam+b\lam$
through back to the left and absorbing
the vertices with spectral parameters $-v-u-\lam-b\lam$ and
$u+v+b\lam$ by the crossing unitary condition
(\ref{fcross-I}), we finally obtain  (\ref{T(u)T(v)})
with $u$ and $v$  exchanged. So we have shown that
$\left[\T^{(p,b)}(v)\; , \T^{(p,q)}(u)\;\right]=0\;$ as required.

{\bigskip\rm\bf\noindent
  Appendix~B: Functional equations and their graphs}
\setcounter{equation}{0}
\renewcommand{\theequation}{B.\arabic{equation}}\newline
In this appendix we prove the functional equations expressed in
Theorem~1. Let us consider $\T^{(q)}_0\T^{(1)}_q$
\be\setlength{\unitlength}{0.0125in}%
\begin{picture}(186,120)(48,675)
\put(105,735){\thicklines\vector( 0, 1){ 60}}
\put(120,735){\thicklines\vector( 0, 1){ 60}}
\put(195,735){\thicklines\vector( 0, 1){ 60}}
\put(180,735){\thicklines\vector( 0, 1){ 60}}
\multiput(138,765)(8.00000,0.00000){4}{
 \makebox(0.4444,0.6667){\tenrm .}}
\put(150,690){\thicklines\vector( 1, 0){  6}}
\put(105,675){\thicklines\vector( 0, 1){ 60}}
\put(120,675){\thicklines\vector( 0, 1){ 60}}
\put(195,675){\thicklines\vector( 0, 1){ 60}}
\put(180,675){\thicklines\vector( 0, 1){ 60}}
\multiput(138,705)(8.00000,0.00000){4}{
 \makebox(0.4444,0.6667){\tenrm .}}
\put(129,720){\thicklines\line(-1, 0){ 39}}
\put( 90,720){\thicklines\line(-1,-1){ 15}}
\put( 75,705){\thicklines\thicklines\line( 1,-1){ 15}}
\put( 90,690){\thicklines\line( 1, 0){120}}
\put(210,690){\thicklines\line( 1, 1){ 15}}
\put(225,705){\thicklines\line(-1, 1){ 15}}
\put(210,720){\thicklines\vector(-1, 0){ 84}}
\put( 75,792){\line( 0,-1){114}}
\put(225,795){\line( 0,-1){114}}
\put(129,780){\line(-1, 0){ 39}}
\put( 90,780){\line(-1,-1){ 15}}
\put( 75,765){\line( 1,-1){ 15}}
\put( 90,750){\line( 1, 0){120}}
\put(210,750){\line( 1, 1){ 15}}
\put(225,765){\line(-1, 1){ 15}}
\put(210,780){\vector(-1, 0){ 84}}
\put(159,750){\vector( 1, 0){  6}}
\put(234,762){\scr$K_-(u\!+\!q\!\lam)$}
\put(22,762){\scr$K_+(u\!+\!q\!\lam)$}
\put(195,771){\scr$u\!\!+\!\!q\!\lam$}
\put(234,702){\scr$K_-^{(q)}(u)$}
\put(30,702){\scr$K_+^{(q)}(u)$}
\put( 99,681){\scr$u$}
\put(195,711){\scr$u\!\!+\!\!q\!\lam\!\!-\!\!\lam$}
\put( 96,747){\small$b$}
\put( 96,717){\small$a$}
\put(168,741){\scr$u\!\!+\!\!q\!\lam$}
\end{picture}\ee
Inserting the operator $Y^+_q$ into the position
$a$ and the identity operator into the position $b$
and then using the unitary condition (\ref{fI}) and
the Yang-Baxter equation (\ref{FYB}), we are
able to obtain
\be\setlength{\unitlength}{0.0125in}%
\begin{picture}(192,105)(-10,525)
\put( 75,630){\line( 0,-1){105}}
\put(225,630){\line( 0,-1){105}}
\put( 75,570){\thicklines\line( 1, 1){ 30}}
\put(105,600){\thicklines\line( 1, 0){ 90}}
\put(195,600){\thicklines\line( 1,-1){ 30}}
\put(225,570){\thicklines\line(-1,-1){ 15}}
\put(210,555){\thicklines\line(-1, 0){120}}
\put( 90,555){\thicklines\line(-1, 1){ 15}}
\put(111,540){\thicklines\vector( 0, 1){ 90}}
\put(126,540){\thicklines\vector( 0, 1){ 90}}
\put(186,540){\thicklines\vector( 0, 1){ 90}}
\put(171,540){\thicklines\vector( 0, 1){ 90}}
\multiput(138,585)(8.00000,0.00000){4}{
 \makebox(0.4444,0.6667){\tenrm .}}
\put(144,555){\vector( 1, 0){  6}}
\put(144,570){\vector( 1, 0){  6}}
\put(144,615){\vector(-1, 0){  3}}
\put(147,600){\vector(-1, 0){  3}}
\put( 75,600){\line( 1,-1){ 30}}
\put(105,570){\line( 1, 0){ 90}}
\put(195,570){\line( 1, 1){ 30}}
\put(225,600){\line(-1, 1){ 15}}
\put(210,615){\line(-1, 0){120}}
\put( 90,615){\line(-1,-1){ 15}}
\put(22,597){\scr$K_+(u\!+\!q\!\lam)$}
\put(30,552){\scr$K_+^{(q)}(u)$}
\put(231,552){\scr$K_-^{(q)}(u)$}
\put(231,597){\scr$K_-(u\!+\!q\!\lam)$}
\put(105,546){\scr$u$}
\put(186,606){\scr$u\!\!+\!\!q\!\lam$}
\put(127,594){\scr$u\!\!+\!\!q\!\lam\!\!-\!\!\lam$}
\put(126,561){\scr$u\!\!+\!\!q\!\lam$}
\put(71,588){\scr$-\!2\!\lam\!\!-\!\!q\!\lam\!\!-\!\!2\!u$}
\put(198,576){\scr$\phantom{+\lam}\!\!\!2\!u\!\!+\!\!q\!\lam$}
\put(102,600){\small$a$}
\put( 96,615){\small$c$}
\put(-52,577){$\disp{1\over
  \tilde{\rho}_{q,1}(2u\!+\!q\lam)}$$\;\times$}
\end{picture}\label{}
\ee
It is easy to see that
the equation can be split into the sum
of the following two parts,
\be
\setlength{\unitlength}{0.0125in}%
\begin{picture}(165,60)(60,735)\thicklines
\put(150,750){\vector( 1, 0){  6}}
\put(105,735){\vector( 0, 1){ 60}}
\put(120,735){\vector( 0, 1){ 60}}
\put(195,735){\vector( 0, 1){ 60}}
\put(180,735){\vector( 0, 1){ 60}}
\multiput(138,765)(8.00000,0.00000){4}{
 \makebox(0.4444,0.6667){\tenrm .}}
\put( 75,789){\thinlines\line( 0,-1){ 45}}
\put(225,786){\thinlines\line( 0,-1){ 45}}
\put(129,780){\line(-1, 0){ 39}}
\put( 90,780){\line(-1,-1){ 15}}
\put( 75,765){\line( 1,-1){ 15}}
\put( 90,750){\line( 1, 0){120}}
\put(210,750){\line( 1, 1){ 15}}
\put(225,765){\line(-1, 1){ 15}}
\put(210,780){\line(-1, 0){ 84}}
\put(140,780){\vector(-1, 0){0}}
\put( 99,743){\scr$u$}
\put(174,743){\scr$u$}
\put(196,773){\scr$u\!\!+\!\!q\!\lam$}
\put(106,773){\scr$u\!\!+\!\!q\!\lam$}
\put(32,764){\scr$K_+^{(q\!+\!1)}(u)$}
\put(230,764){\scr$K_-^{(q\!+\!1)}(u)$}
\end{picture} \label{B1}
\ee
and
\be
\setlength{\unitlength}{0.0125in}%
\begin{picture}(165,60)(40,735)\thicklines
\put(150,750){\vector( 1, 0){  6}}
\put(105,735){\vector( 0, 1){ 60}}
\put(120,735){\vector( 0, 1){ 60}}
\put(195,735){\vector( 0, 1){ 60}}
\put(180,735){\vector( 0, 1){ 60}}
\multiput(138,765)(8.00000,0.00000){4}{
 \makebox(0.4444,0.6667){\tenrm .}}
\put( 75,789){\thinlines\line( 0,-1){ 45}}
\put(225,786){\thinlines\line( 0,-1){ 45}}
\put(129,780){\line(-1, 0){ 39}}
\put( 90,780){\line(-1,-1){ 15}}
\put( 75,765){\line( 1,-1){ 15}}
\put( 90,750){\line( 1, 0){120}}
\put(210,750){\line( 1, 1){ 15}}
\put(225,765){\line(-1, 1){ 15}}
\put(210,780){\line(-1, 0){ 84}}
\put(140,780){\vector(-1, 0){0}}
\put( 99,743){\scr$u$}
\put(174,743){\scr$u$}
\put(196,773){\scr$u\!\!+\!\!q\!\lam\!\!-\!\!2\!\lam$}
\put(106,773){\scr$u\!\!+\!\!q\!\lam\!\!-\!\!2\!\lam$}
\put(-28,764){$\ol{f}^{p}_{q-1}(u)\;
 \times$\scr$\hs{0.1}\biggl(\;K_+^{(q\!-\!1)}(u)$}
\put(230,764){\scr$K_-^{(q\!-\!1)}(u)\biggl)$}
\end{picture}\label{B2}
\ee
where the $u$-dependent function $\ol{f}^p_{q-1}(u)$ is
generated from the  antisymmetric fusion of the model.
The underlying models are su($2$) type and the antisymmetric
fusion gives only one independent nonzero element for the
$R$ or $K$ matrices. Therefore $\ol{f}^p(u)$ can be factorized as
\be
\ol{f}^p_q(u)&=& \omega_-(u+q\lam-\lam)\omega_+(u+q\lam-\lam)
    \phi_+^p(u+q\lam-\lam) \no \\
 &&\h\times\phi_-^p(u+q\lam-\lam)
 I_+(u+q\lam-\lam|q)I_-(u+q\lam-\lam|q)\;.
\ee
For the six- and eight-vertex models we have shown
\be
I_+(u|q)I_-(u|q)=I_+(u|1)I_-(u|1)=\tilde{\rho}_{1,1}^{-1}(2u+\lam)\;.
\ee
Therefore $f^p_q(u)$ depends on $q$ only through the
spectra parameter shift $u+q\lam-\lam$. We suppress
the subscript $q$ and define $f^p(u)=\ol{f}^p_q(u-q\lam+\lam)$.
Graphically  $f^p(u)$ is given by
\be
\setlength{\unitlength}{0.0090in}%
\begin{picture}(309,120)(0,687)
\put(105,792){\line( 0,-1){ 90}}
\put(105,762){\line( 1, 1){ 30}}
\put(135,792){\line( 1,-1){ 30}}
\put(165,792){\line(-1,-1){ 30}}
\put(135,762){\line(-1,-1){ 30}}
\put(105,732){\line( 1,-1){ 30}}
\put(135,702){\line( 1, 0){ 81}}
\put(105,762){\line( 1,-1){ 30}}
\put(135,732){\line( 1, 0){ 27}}
\put(162,732){\line( 1, 0){ 51}}
\put(165,762){\line( 1, 0){ 48}}
\put(165,792){\line( 1, 0){ 48}}
\put(150,777){\circle{16}}
\put(210,762){\line( 1, 0){ 48}}
\put(180,687){\thicklines\vector( 0, 1){120}}
\put(201,687){\thicklines\vector( 0, 1){120}}
\put(294,687){\thicklines\vector( 0, 1){120}}
\put(273,687){\thicklines\vector( 0, 1){120}}
\put(237,762){\vector(-1, 0){  3}}
\put(240,792){\vector(-1, 0){  3}}
\multiput(222,747)(10.00000,0.00000){4}{
 \makebox(0.4444,0.6667){\tenrm .}}
\put(366,792){\line( 0,-1){ 93}}
\put(225,762){\line( 1, 0){111}}
\put(336,762){\line( 1,-1){ 30}}
\put(366,732){\line(-1,-1){ 30}}
\put(336,702){\line(-1, 0){129}}
\put(231,702){\vector( 1, 0){  6}}
\put(228,732){\vector( 1, 0){  9}}
\put(210,792){\line( 1, 0){126}}
\put(336,792){\line( 1,-1){ 30}}
\put(366,762){\line(-1,-1){ 30}}
\put(336,732){\line(-1, 0){123}}
\put(338,747){\scr$2\!u\!\!+\!\!\lam$}
\put(160,720){\scr$u\!\!+\!\!\lam$}
\put(274,753){\scr$u$}
\put(274,783){\scr$u\!\!+\!\!\lam$}
\put(174,693){\scr$u$}
\put(369,726){\scr$K_-(u)$}
\put(369,759){\scr$K_-(u+\lam)$}
\put( 63,729){\scr$K_+(u)$}
\put(-53,741){$\disp{1\over\tilde{\rho}_{1,1}(2u\!+\!\lam)}$}
\put(40,756){\scr$K_+(u+\lam)$}
\put(108,750){\scr$-\!2\!u\!\!-\!\!3\!\lam$}
\end{picture}
\label{fp}\ee
or
\be
{f}^p(u)&=& \omega_-(u)\omega_+(u)
    \phi_+^p(u)\phi_-^p(u)
 I_+(u|q)I_-(u|q) \label{B3}
\ee
where
\be
\setlength{\unitlength}{0.0105in}%
\begin{picture}(397,102)(33,696)
\put(333,714){\circle{14}}
\put(378,783){\line( 0,-1){ 78}}
\put(333,789){\line( 1, 0){ 15}}
\put(348,789){\line( 1,-1){ 30}}
\put(378,759){\line(-1,-1){ 60}}
\put(333,759){\line( 1, 0){ 15}}
\put(348,759){\line( 1,-1){ 30}}
\put(378,729){\line(-1,-1){ 30}}
\put(348,699){\line(-1, 1){ 30}}
\put(342,759){\vector(-1, 0){  6}}
\put(339,789){\vector(-1, 0){  3}}
\put(312,726){\scr$i$}
\put(309,696){\scr$k$}
\put(327,786){\scr$j$}
\put(327,753){\scr$l$}
\put(381,753){\scr$K_-(u\!\!+\!\!\lam)$}
\put(381,726){\scr$K_-(u)$}
\put(351,735){\scr$2u\!\!+\!\!\lam$}
\put(265,738){$\omega_-(u) =$}
\put(189,774){\circle{14}}
\put(144,783){\line( 0,-1){ 78}}
\put(144,759){\line( 1, 1){ 30}}
\put(174,789){\line( 1,-1){ 30}}
\put(144,759){\line( 1,-1){ 30}}
\put(174,729){\line( 1, 0){ 15}}
\put(189,729){\line(-1, 0){  3}}
\put(189,699){\line(-1, 0){ 15}}
\put(174,699){\line(-1, 1){ 30}}
\put(144,729){\line( 1, 1){ 60}}
\put(180,729){\vector( 1, 0){  6}}
\put(183,699){\vector( 1, 0){  3}}
\put( 96,756){\scr$K_+(u\!\!+\!\!\lam)$}
\put(111,726){\scr$K_+(u)$}
\put(147,741){\scr$-\!2\!u\!\!-\!\!3\!\lam$}
\put(207,783){\scr$i$}
\put(207,753){\scr$k$}
\put(192,696){\scr$l$}
\put(192,726){\scr$j$}
\put(43,738){$\omega_+(u) =$}
\end{picture}
\ee
\be\setlength{\unitlength}{0.0105in}%
\begin{picture}(405,54)(1,666)
\multiput(147,690)(6.00000,0.00000){4}{
 \makebox(0.4444,0.6667){\tenrm .}}
\put(126,669){\thicklines\vector( 0, 1){ 51}}
\put(141,669){\thicklines\vector( 0, 1){ 51}}
\put(180,669){\thicklines\vector( 0, 1){ 51}}
\put(195,669){\thicklines\vector( 0, 1){ 51}}
\put(370,711){\thicklines\vector( 0,-1){ 51}}
\put(365,713){\scr$q\!\!-\!\!1$}
\multiput( 81,675)(240,0){2}{\line( 1, 1){ 30}}
\multiput( 81,705)(240,0){2}{\line( 1,-1){ 30}}
\put(111,705){\vector( 1, 0){ 96}}
\put(111,675){\vector( 1, 0){ 96}}
\put(351,705){\vector( 1, 0){ 66}}
\put(351,675){\vector( 1, 0){ 66}}
\put(171,666){\scr$u$}
\put(165,696){\scr$u\!\!+\!\!\lam$}
\put(372,696){\scr$(\!q\!\!-\!\!4\!)\!\lam\!\!-\!\!2\!u$}
\put(372,666){\scr$(\!q\!\!-\!\!3\!)\!\lam\!\!-\!\!2\!u$}
\put(250,687){$I_+(u|q)=$}
\put(-24,687){$\phi_+^p(\!u\!\!-\!\!\phalf\lam
  \!\!+\!\!\half\!\lam\!)\!=$}
\multiput( 96,690)(240,0){2}{\circle{14}}
\multiput(207,705)(210,0){2}{\scr$j$}
\multiput(207,675)(210,0){2}{\scr$l$}
\multiput(76,705)(240,0){2}{\scr$i$}
\multiput(76,675)(240,0){2}{\scr$k$}
\end{picture}\ee
\be
\setlength{\unitlength}{0.0100in}%
\begin{picture}(111,127)(145,672)
\multiput( 96,690)(200,20){2}{\circle{16}}
\put( 75,720){\thicklines\vector( 1, 0){ 51}}
\put( 75,735){\thicklines\vector( 1, 0){ 51}}
\put( 75,774){\thicklines\vector( 1, 0){ 51}}
\put( 75,789){\thicklines\vector( 1, 0){ 51}}
\multiput( 81,675)(200,20){2}{\line( 1, 1){ 30}}
\multiput(111,675)(200,20){2}{\line(-1, 1){ 30}}
\put(111,705){\vector( 0, 1){ 96}}\put(81,705){\vector( 0, 1){ 96}}
\put(311,725){\vector( 0, 1){ 66}}\put(281,725){\vector( 0, 1){66}}
\put(326,754){\thicklines\vector(-1,0){59}}
\multiput( 96,741)(0.00000,6.00000){4}{
 \makebox(0.4444,0.6667){\tenrm .}}
\put(302,747){\scr$2\!u\!\!-\!\!q\!\lam\!\!+\!\!\lam$}
\put(252,758){\scr$2\!u\!\!-\!\!q\!\lam\!\!+\!\!2\!\lam$}
\put( 75,738){\scr$u$}
\put(114,738){\scr$u\!\!+\!\!\lam$}
\multiput( 81,804)(200,-10){2}{\scr$j$}
\multiput(117,672)(200,20){2}{\scr$k$}
\multiput( 72,672)(200,20){2}{\scr$i$}
\multiput(111,801)(200,-10){2}{\scr$l$}
\put(-36,741){$\disp{\phi_-^p
 (\!u\!\!+\!\!\phalf\lam\!\!-\!\!\half\!\lam\!)\!=}$}
\put(330,772){\scr$q\!\!-\!\!1$}\put(330,772){\vector(-1,-1){15}}
\put(190,741){$\disp{I_-(u|q)=}$}
\put(345,735){$\disp{\times\tilde{
  \rho}_{q,1}^{-1}(2\!u\!\!-\!\!q\!\lam\!\!+\!\!2\!\lam)}$}
\put(345,755){$\disp{\times\tilde{
 \rho}_{q\!-\!1,1}^{-1}(2\!u\!\!-\!\!q\!\lam\!\!+\!\!\lam)}$}
\end{picture}
\ee
where the cross with a circle plays the role of antisymmetric
operator,
$$\setlength{\unitlength}{0.0100in}%
\begin{picture}(30,30)(115,660)
\put( 90,675){\circle{16}}
\put( 105,660){\vector( -1,1){ 30}}
\put(105,690){\vector(-1,-1){ 30}}
\put( 69,684){\small$i$}
\put(125,669){$=\half(\delta_{ik}
 \delta_{jl}-\delta_{jk}\delta_{il})\;.$}
\put(108,654){\small$l$}
\put( 69,654){\small$j$}
\put(108,684){\small$k$}
\end{picture}$$
The theorem~1 can be seen from (\ref{B1}), (\ref{B2}) and (\ref{B3}).

{\bigskip\rm\bf\noindent  Appendix~C: Intertwining
  relations and their graphs}
\setcounter{equation}{0}
\renewcommand{\theequation}{C.\arabic{equation}}\newline
Acoording to the graphs presented in Appendix~B and Appendix~C the
intertwining relations between the vertex and face models (\ref{RW})
can be represented as
\be\setlength{\unitlength}{0.0105in}%
\begin{picture}(161,90)(0,0)
\put(-30,0){
\begin{picture}(81,90)(87,609)
\put(120,675){\circle*{4}}
\put(150,645){\circle*{4}}
\put( 90,690){\line( 1, 0){ 60}}
\put(165,675){\line( 0,-1){ 60}}
\put(120,690){\vector( 1, 0){  3}}
\put(165,654){\vector( 0,-1){  9}}
\put(120,687){\line( 0,-1){ 72}}
\put( 90,645){\line( 1, 0){ 75}}
\put(135,645){\vector( 1, 0){  6}}
\put(120,660){\vector( 0, 1){  6}}
\put(120,675){\vector( 0, 1){ 15}}
\put(150,645){\vector( 1, 0){ 15}}
\put( 87,690){\small$d$}
\put(153,687){\small$c$}
\put(168,672){\small$c$}
\put(168,609){\small$b$}
\put(123,612){\small$j$}
\put( 90,633){\small$i$}
\put(123,648){\small$u\!-\!v$}
\put(108,693){\small$u$}
\put(168,630){\small$v$}
\end{picture}}
\put(97,35){$=$}
\put(130,-14){
\begin{picture}(96,99)(210,600)
\put(240,615){\circle*{4}}
\put(240,630){\circle*{4}}
\put(225,630){\circle*{4}}
\put(240,690){\line( 1, 0){ 60}}
\put(300,690){\line( 0,-1){ 60}}
\put(300,630){\line(-1, 0){ 60}}
\put(240,630){\line( 0, 1){ 60}}
\put(240,615){\line( 1, 0){ 60}}
\put(225,690){\line( 0,-1){ 60}}
\put(270,690){\vector( 1, 0){  3}}
\put(267,630){\vector( 1, 0){  3}}
\put(240,666){\vector( 0,-1){  9}}
\put(300,666){\vector( 0,-1){  9}}
\put(225,666){\vector( 0,-1){  9}}
\put(270,615){\vector( 1, 0){  3}}
\put(210,660){\vector( 1, 0){ 15}}
\put(270,600){\vector( 0, 1){ 15}}
\put(231,690){\small$d$}
\put(306,687){\small$c$}
\put(306,621){\small$b$}
\put(210,648){\small$i$}
\put(273,600){\small$j$}
\put(256,657){\small$u\!-\!v$}
\put(219,669){\small$v$}
\put(288,606){\small$u$}
\end{picture}}\end{picture}\ee
where the solid dots mean  summation (as do
other solid dots in this section).
The Star-triangle relation (Yang-Baxter equation)
of these face weights is graphically
\def\YBR#1#2#3#4#5#6#7#8#9{
\setlength{\unitlength}{0.010in}%
\begin{picture}(300,100)(-10,-10)
  \multiput(60,60)(0,-60){2}{\line(1,0){40}}
 \multiput(80,60)(0,-60){2}{\vector(1,0){0}}
  \multiput(60,60)(60,-30){2}{\line(-2,-3){20}}
 \multiput(60,60)(60,-30){2}{\vector(-2,-3){10}}
  \multiput(40,30)(60,30){2}{\line(2,-3){20}}
 \multiput(40,30)(60,30){2}{\vector(2,-3){10}}
 \put(40,30){\line(1,0){40}} \put(40,30){\vector(1,0){20}}
 \put(100,60){\line(-2,-3){20}}\put(100,60){\vector(-2,-3){10}}
 \put(100,0){\line(-2,3){20}}\put(90,15){\vector(2,-3){0}}
 \put(80,30){\circle*{3}}
 \put(140,30){$=$}
  \multiput(190,60)(0,-60){2}{\line(1,0){40}}
 \multiput(216,60)(0,-60){2}{\vector(1,0){0}}
  \multiput(190,60)(60,-30){2}{\line(-2,-3){20}}
 \multiput(190,60)(60,-30){2}{\vector(-2,-3){10}}
  \multiput(170,30)(60,30){2}{\line(2,-3){20}}
 \multiput(170,30)(60,30){2}{\vector(2,-3){10}}
 \put(190,60){\line(2,-3){20}}\put(190,60){\vector(2,-3){10}}
    \put(190,0){\line(2,3){20}} \put(198,12){\vector(-2,-3){0}}
 \put(250,30){\line(-1,0){40}}\put(233,30){\vector(1,0){0}}
 \put(210,30){\circle*{3}}
  \multiput(57,-7)(130,0){2}{\scr $#1$}
  \multiput(100,-7)(130,0){2}{\scr $#2$}
  \multiput(120,28)(130,0){2}{\scr $#3$}
  \multiput(101,63)(130,0){2}{\scr $#4$}
  \multiput(57,63)(130,0){2}{\scr $#5$}
  \multiput(35,28)(130,0){2}{\scr $#6$}
 \put(68,43){\scr $#7$} %u
 \put(90,28){\scr $#8$} %u-v
 \put(68,13){\scr $#9$} %v
 \put(218,43){\scr $#9$} %v
 \put(180,28){\scr $#8$} %u-v
 \put(218,13){\scr $#7$} %u}
\end{picture}}
\be
 \YBR abcdefv{v\!-\!u}u \label{YBR}
\ee
For the boundaries the correspondence (\ref{KK}) are given by
\be
\setlength{\unitlength}{0.0095in}%
\begin{picture}(142,96)(0,0)
\put(-30,0){
\begin{picture}(72,96)(63,705)
\put(105,765){\circle*{4}}
\put(105,705){\line( 1, 1){ 30}}
\put(135,735){\vector(-1, 1){ 45}}
\put(120,720){\vector( 1, 1){0}}
\put(111,801){\line(-1,-1){ 45}}
\put( 87,777){\vector(1,1){0}}
\put( 63,747){\small$b$}
\put(114,792){\small$a$}
\put( 99,708){\small$i$}
\put( 81,759){\small$u$}
\put(129,744){\small$u$}
\end{picture}}
\put(90,30){$=$}
\put(140,0){
\begin{picture}(49,100)(168,698)
\put(213,732){\circle*{4}}
\put(213,702){\circle*{4}}
\put(213,715){\small$c$}
\put(198,777){\line(1,1){15}}
\put(183,762){\vector(1,1){15}}
\put(183,762){\line( 1,-1){ 30}}
\put(213,732){\line( 0, 1){ 60}}
\put(198,747){\vector( 1,-1){0}}
\put(168,747){\vector( 1,-1){ 24}}
\put(183,732){\line( 1,-1){ 30}}
\put(173,710){\vector( 1, 1){16}}
\put(174,750){\small$b$}
\put(171,717){\small$i$}
\put(204,792){\small$a$}
\put(185,705){\small$-\!u$}
\put(198,759){\small$u$}
\end{picture}}
\end{picture}\ee
and
\be
\setlength{\unitlength}{0.0095in}%
\begin{picture}(133,93)(0,0)
\put(0,0){
\begin{picture}(63,93)(120,687)
\put(150,780){\vector(-1,-1){ 15}}
\put(135,765){\line(-1,-1){ 15}}
\put(120,750){\vector( 1,-1){ 45}}
\put(150,720){\circle*{4}}
\put(183,723){\vector(-1,-1){ 18}}
\put(147,687){\line( 1, 1){ 18}}
\put(150,768){\small$i$}
\put(177,729){\small$b$}
\put(144,690){\small$a$}
\put(126,744){\small$u$}
\put(159,690){\small$-\!u\!-\!\lam$}
\end{picture}}
\put(90,40){$=\disp{S(c)\over S(b)}$}
\put(145,0){
\begin{picture}(49,100)(134,696)
\put(138,762){\circle*{4}}
\put(138,792){\circle*{4}}
\put(138,770){\small$c$}
\put(168,732){\vector(-1, 1){ 15}}
\put(153,747){\line(-1, 1){ 15}}
\put(138,762){\line( 0,-1){ 60}}
\put(138,702){\line( 1, 1){ 15}}
\put(138,792){\line( 1,-1){ 45}}
\put(175,776){\small$i$}
\put(174,783){\vector(-1,-1){ 15}}
\put(159,771){\vector(-1, 1){0}}
\put(168,732){\vector(-1,-1){ 15}}
\put(174,738){\small$b$}
\put(144,696){\small$a$}
\put(146,784){\small$u\!\!+\!\!\lam$}
\put(141,726){\small$u$}
\end{picture}}
\end{picture}\ee
The unitary conditions of the face weights (\ref{Wunit}) is
\be
\setlength{\unitlength}{0.009in}%
\begin{picture}(135,84)(126,723)
\put(135,765){\line( 1, 1){ 30}}
\put(165,795){\line( 1,-1){ 30}}
\put(195,765){\line( 1,-1){ 30}}
\put(225,735){\line( 1, 1){ 30}}
\put(255,765){\line(-1, 1){ 30}}
\put(225,795){\line(-1,-1){ 60}}
\put(165,735){\line(-1, 1){ 30}}
\put(195,765){\circle*{4}}
\put(193,770){\scr$c$}
\put(207,777){\vector(-1,-1){0}}
\put(183,777){\vector( 1,-1){0}}
\put(156,744){\vector( 1,-1){0}}
\put(213,747){\vector( 1,-1){0}}
\put(243,777){\vector( 1,-1){0}}
\put(237,747){\vector(-1,-1){0}}
\put(180,750){\vector(-1,-1){0}}
\put(150,780){\vector(-1,-1){0}}
\put(159,759){\small$u$}
\put(219,759){\small$-u$}
\put(162,798){\small$d$}
\put(222,798){\small$d$}
\put(126,762){\small$a$}
\put(258,762){\small$a'\;\;\;=\;\mbox{$\rho(u)\delta_{aa'}$}$}
\put(225,723){\small$b$}
\put(165,723){\small$b$}
\end{picture}
\ee
and the crossing unitary condition (\ref{Wcrossunit}) is
\be
\setlength{\unitlength}{0.009in}%
\begin{picture}(150,84)(105,708)
\put( 84,750){\line( 1, 1){ 30}}
\put(114,780){\line( 1,-1){ 30}}
\put(144,750){\line( 1,-1){ 30}}
\put(174,720){\line( 1, 1){ 30}}
\put(204,750){\line(-1, 1){ 30}}
\put(174,780){\line(-1,-1){ 60}}
\put(144,750){\circle*{4}}
\put(142,756){\scr$c$}
\put(114,720){\line(-1, 1){ 30}}
\put(162,732){\vector( 1,-1){0}}
\put(192,762){\vector( 1,-1){0}}
\put(129,735){\vector(-1,-1){0}}
\put( 99,765){\vector(-1,-1){0}}
\put(162,768){\vector( 1, 1){0}}
\put(195,741){\vector( 1, 1){0}}
\put(123,771){\vector(-1, 1){0}}
\put( 96,738){\vector(-1, 1){0}}
\put(111,783){\small$d$}
\put(171,783){\small$d$}
\put( 75,747){\small$a$}
\put(208,747){\small$a'$}
\put(174,708){\small$b$}
\put(114,708){\small$b$}
\put(158,744){\small$u\!-\!\lam$}
\put(95,744){\small$-\!u\!-\!\lam$}
\put(225,747){$\disp{{S(a)S(c)\over S(b)S(d)}\;\;\;=\;
          \rho(u)\delta_{aa'}}$}
\end{picture}
\ee

The reflection equations for the SOS model
(\ref{WK+}) and (\ref{WK-}) can be presented
graphically by
\be
\setlength{\unitlength}{0.0095in}%
\begin{picture}(199,135)(-10,0)
\put(0,0){\begin{picture}(109,135)(99,681)
\put(165,780){\circle*{4}}
\put(135,750){\line(-1, 1){ 30}}
\put(105,780){\line( 1, 1){ 30}}
\put(135,810){\line( 1,-1){ 60}}
\put(195,750){\line( 0,-1){ 60}}
\put(195,690){\line(-1, 1){ 60}}
\put(135,750){\line( 1, 1){ 60}}
\put(195,810){\line( 0,-1){ 60}}
\put(195,750){\line(-1,-1){ 30}}
\put(120,765){\vector(-1, 1){0}}
\put(147,798){\vector(-1, 1){0}}
\put(153,732){\vector( 1,-1){0}}
\put(186,759){\vector( 1,-1){0}}
\put(186,699){\vector( 1,-1){0}}
\put(153,768){\vector( 1, 1){0}}
\put(123,798){\vector( 1, 1){0}}
\put(183,738){\vector( 1, 1){0}}
\put(183,798){\vector( 1, 1){0}}
\put(126,741){\small$a$}
\put( 99,777){\small$b$}
\put(135,810){\small$c$}
\put(192,810){\small$c$}
\put(159,708){\small$d$}
\put(198,744){\small$f$}\put(195,750){\circle*{4}}
\put(192,681){\small$e$}
\put(183,777){\small$u$}
\put(183,714){\small$v$}
\put(150,747){\small$u\!+\!v$}
\put(123,777){\small$u\!-\!v$}
\end{picture}}
\put(120,70){$=$}
\put(140,0){
\begin{picture}(99,135)(132,678)
\put(198,720){\circle*{4}}
\put(168,750){\line(-1,-1){ 30}}
\put(138,720){\line( 1,-1){ 30}}
\put(168,690){\line( 1, 1){ 60}}
\put(228,750){\line( 0,-1){ 60}}
\put(228,690){\line(-1, 1){ 60}}
\put(168,750){\line( 1, 1){ 60}}
\put(228,810){\line( 0,-1){ 60}}
\put(228,750){\line(-1, 1){ 30}}
\put(159,699){\vector( 1,-1){0}}
\put(186,732){\vector( 1,-1){0}}
\put(219,759){\vector( 1,-1){0}}
\put(186,768){\vector( 1, 1){0}}
\put(216,798){\vector( 1, 1){0}}
\put(216,738){\vector( 1, 1){0}}
\put(213,705){\vector( 1,-1){0}}
\put(180,702){\vector(-1,-1){0}}
\put(150,732){\vector(-1,-1){0}}
\put(228,678){\small$e$}
\put(168,678){\small$e$}
\put(162,750){\small$a$}
\put(192,780){\small$b$}
\put(231,807){\small$c$}
\put(231,744){\small$f$}\put(228,750){\circle*{4}}
\put(132,714){\small$d$}
\put(186,747){\small$u\!+\!v$}
\put(156,717){\small$u\!-\!v$}
\put(216,717){\small$u$}
\put(216,777){\small$v$}
\end{picture}}
\end{picture}
\ee
and
\be
\setlength{\unitlength}{0.0095in}%
\begin{picture}(96,135)(40,0)
\put(0,0){
\begin{picture}(96,135)(102,678)
\put(138,777){\circle*{4}\small$g$}
\put(168,747){\line( 1, 1){ 30}}
\put(198,777){\line(-1, 1){ 30}}
\put(168,807){\line(-1,-1){ 60}}
\put(108,747){\line( 0,-1){ 60}}
\put(108,687){\line( 1, 1){ 60}}
\put(168,747){\line(-1, 1){ 60}}
\put(108,807){\line( 0,-1){ 60}}
\put(108,747){\line( 1,-1){ 30}}
\put(123,792){\vector(-1, 1){0}}
\put(120,735){\vector(-1, 1){0}}
\put(150,765){\vector(-1, 1){0}}
\put(180,795){\vector(-1, 1){0}}
\put(183,762){\vector( 1, 1){0}}
\put(156,795){\vector( 1, 1){0}}
\put(150,729){\vector(-1,-1){0}}
\put(120,759){\vector(-1,-1){0}}
\put(120,699){\vector(-1,-1){0}}
\put(198,774){\small$b$}
\put(96,744){\small$f$}\put(108,747){\circle*{4}}
\put(105,807){\small$c$}
\put(168,807){\small$c$}
\put(105,678){\small$e$}
\put(168,741){\small$a$}
\put(138,708){\small$d$}
\put(120,771){\small$u$}
\put(117,714){\small$v$}
\put(153,774){\small$v\!-\!u$}
\put(114,744){\scr$-\!2\!\lam\!\!-\!\!u\!\!-\!\!v$}
\put(20,750){$\disp{S(f)S(a)\over S(d)S(g)}$}
\end{picture}}
\put(120,70){$=$ $\disp{S(f)S(a)\over S(b)S(g)}$}
\put(210,0){\begin{picture}(96,138)(110,684)
\put(156,723){\circle*{4}\small$g$}
\put(141,798){\vector(-1, 1){0}}
\put(141,738){\vector(-1, 1){0}}
\put(168,771){\vector(-1, 1){0}}
\put(177,702){\vector( 1,-1){0}}
\put(204,735){\vector( 1,-1){0}}
\put(138,705){\vector(-1,-1){0}}
\put(138,765){\vector(-1,-1){0}}
\put(168,735){\vector(-1,-1){0}}
\put(198,705){\vector(-1,-1){0}}
\put(186,753){\line( 1,-1){ 30}}
\put(216,723){\line(-1,-1){ 30}}
\put(186,693){\line(-1, 1){ 60}}
\put(126,753){\line( 0,-1){ 60}}
\put(126,693){\line( 1, 1){ 60}}
\put(186,753){\line(-1, 1){ 60}}
\put(126,813){\line( 0,-1){ 60}}
\put(126,753){\line( 1, 1){ 30}}
\put(186,753){\small$a$}
\put(159,783){\small$b$}
\put(216,723){\small$d$}
\put(113,746){\small$f$}\put(126,753){\circle*{4}}
\put(126,816){\small$c$}
\put(126,684){\small$e$}
\put(183,684){\small$e$}
\put(135,780){\small$v$}
\put(135,720){\small$u$}
\put(134,747){\scr$-\!2\!\lam\!\!-\!\!u\!\!-\!\!v$}
\put(171,720){\small$v\!-\!u$}
\end{picture}}
\end{picture}
\ee

The transfer matrix of the SOS model with the open
boundaries (\ref{V}) is presented graphically by
\be
\setlength{\unitlength}{0.0095in}%
\begin{picture}(328,85)(-36,731)
\put( 90,780){\circle*{4}}
\put(105,780){\circle*{4}}
\put(315,780){\circle*{4}}
\put(330,780){\circle*{4}}
\put(105,810){\line( 1, 0){210}}
\put(315,810){\line( 0,-1){ 60}}
\put(315,750){\line(-1, 0){210}}
\put(105,750){\line( 0, 1){ 60}}
\put(105,780){\line( 1, 0){210}}
\put(135,810){\line( 0,-1){ 60}}
\put(165,810){\line( 0,-1){ 60}}
\put(285,810){\line( 0,-1){ 60}}
\put(255,810){\line( 0,-1){ 60}}
\multiput(195,765)(7.33333,0.00000){5}{\line( 1, 0){  3.667}}
\multiput(195,795)(7.33333,0.00000){5}{\line( 1, 0){  3.667}}
\put( 60,810){\line( 1,-1){ 30}}
\put( 90,780){\line(-1,-1){ 30}}
\put(360,810){\line(-1,-1){ 30}}
\put(330,780){\line( 1,-1){ 30}}
\put( 75,795){\line(-1,-1){ 15}}
\put( 60,780){\line( 1,-1){ 15}}
\put(345,795){\line( 1,-1){ 15}}
\put(360,780){\line(-1,-1){ 15}}
\put( 63,783){\vector(-1,-1){0}}
\put(354,774){\vector( 1, 1){0}}
\put( 75,795){\vector(-1, 1){0}}
\put( 72,762){\vector(-1,-1){0}}
\put(105,789){\vector( 0, 1){  6}}
\put(105,771){\vector( 0,-1){  9}}
\put(117,780){\vector( 1, 0){  6}}
\put(135,771){\vector( 0,-1){  9}}
\put(165,771){\vector( 0,-1){  9}}
\put(255,768){\vector( 0,-1){  9}}
\put(285,768){\vector( 0,-1){  9}}
\put(315,768){\vector( 0,-1){  9}}
\put(135,792){\vector( 0, 1){  6}}
\put(165,792){\vector( 0, 1){  6}}
\put(255,792){\vector( 0, 1){  6}}
\put(285,792){\vector( 0, 1){  6}}
\put(315,792){\vector( 0, 1){  6}}
\put(120,750){\vector( 1, 0){  6}}
\put(120,810){\vector( 1, 0){  6}}
\put(147,810){\vector( 1, 0){  6}}
\put(150,780){\vector( 1, 0){  6}}
\put(147,750){\vector( 1, 0){  6}}
\put(207,780){\vector( 1, 0){  6}}
\put(207,810){\vector( 1, 0){  6}}
\put(210,750){\vector( 1, 0){  6}}
\put(270,750){\vector( 1, 0){  6}}
\put(270,780){\vector( 1, 0){  6}}
\put(270,810){\vector( 1, 0){  6}}
\put(300,810){\vector( 1, 0){  6}}
\put(297,780){\vector( 1, 0){  6}}
\put(300,750){\vector( 1, 0){  6}}
\put(345,795){\vector( 1, 1){0}}
\put(348,762){\vector( 1,-1){0}}
\put(96,810){\small$a_0$}
\put(321,807){\small$a_N$}
\put(66,810){\small$a_0$}
\put(351,807){\small$a_N$}
\put(66,741){\small$b_0$}
\put(90,741){\small$b_0$}
\put(321,738){\small$b_N$}
\put(351,735){\small$b_N$}
\put(107,789){\scr$u\!\!-\!\!v_1$}
\put(107,759){\scr$u\!\!+\!\!v_1$}
\put(140,759){\scr$u\!\!+\!\!v_2$}
%\put(264,759){\scr$u\!\!+\!\!v_{N-1}$}
\put(287,759){\scr$u\!\!+\!\!v_{\!N}$}
\put(140,789){\scr$u\!\!-\!\!v_2$}
\put(287,789){\scr$u\!\!-\!\!v_{\!N}$}
%\put(264,789){\scr$u\!\!-\!\!v$}
\put(363,774){\small$K_-(u)$}
\put(-95,774){$\langle\mbox{\boldmath $a$}|
  {\bf V}(u)|\mbox{\boldmath $b$}\rangle=$\small$K_+(u)$}
\end{picture}
\ee
where the boundaries are given by
\be
\setlength{\unitlength}{0.0095in}%
\begin{picture}(72,75)(-154,729)
\put( 93,798){\line( 1,-1){ 30}}
\put(123,768){\line(-1,-1){ 30}}
\put(108,783){\line(-1,-1){ 15}}
\put( 93,768){\line( 1,-1){ 15}}
\put( 96,771){\vector(-1,-1){0}}
\put(108,783){\vector(-1, 1){0}}
\put(105,750){\vector(-1,-1){0}}
\put( 99,798){\small$a$}
\put(123,762){\small$b$}
\put( 99,729){\small$c$}
\put(44,762){\small$K_+(u)$}
\put(-370,762){$
\Kp bacu :=\disp{S(b)\over S(a)}
\varphi_{b,a}^{-1}(u+\lam)\stackrel{t}{K}_+(u)\varphi_{b,c}(-u-\lam)=$}
\end{picture}
\ee
and
\be
\setlength{\unitlength}{0.0095in}%
\begin{picture}(58,81)(-30,726)
\put(162,801){\line(-1,-1){ 30}}
\put(132,771){\line( 1,-1){ 30}}
\put(147,786){\line( 1,-1){ 15}}
\put(162,771){\line(-1,-1){ 15}}
\put(156,765){\vector( 1, 1){0}}
\put(147,786){\vector( 1, 1){0}}
\put(150,753){\vector( 1,-1){0}}
\put(153,798){\small$a$}
\put(123,765){\small$b$}
\put(153,726){\small$c$}
\put(165,765){\small$K_-(u)$}
\put(-185,765){$\Km bacu :=
\varphi_{b,c}^{-1}(-u)K_-(u)\varphi_{b,a}(u)=$}
\end{picture}
\ee

Let us consider $\V(u)\V(u+\lam)$ graphically,
\be\setlength{\unitlength}{0.009in}%
\begin{picture}(217,120)(56,690)
\put( 60,750){\circle*{4}}
\put( 90,750){\circle*{4}}
\put(120,750){\circle*{4}}
\put(150,750){\circle*{4}}
\put(210,750){\circle*{4}}
\put(240,750){\circle*{4}}
\put(270,750){\circle*{4}}
\put( 90,810){\line( 1, 0){150}}
\put(240,810){\line( 0,-1){ 60}}
\put(240,750){\line( 0,-1){ 60}}
\put(240,690){\line(-1, 0){150}}
\put( 90,690){\line( 0, 1){120}}
\put( 90,780){\line( 1, 0){150}}
\put( 90,750){\line( 1, 0){150}}
\put( 90,720){\line( 1, 0){150}}
\put(120,810){\line( 0,-1){120}}
\put(150,810){\line( 0,-1){120}}
\put(210,810){\line( 0,-1){120}}
\put(270,810){\line(-1,-1){ 30}}
\put(240,780){\line( 1,-1){ 30}}
\put(270,750){\line(-1,-1){ 30}}
\put(240,720){\line( 1,-1){ 30}}
\put( 60,810){\line( 1,-1){ 30}}
\put( 90,780){\line(-1,-1){ 30}}
\put( 60,750){\line( 1,-1){ 30}}
\put( 90,720){\line(-1,-1){ 30}}
\put(270,810){\line( 0,-1){120}}
\put( 60,810){\line( 0,-1){120}}
\multiput( 90,792)(30,0){3}{\vector( 0, 1){  6}}
\multiput(210,789)(30,0){2}{\vector( 0, 1){  6}}
\multiput( 90,765)(30,0){3}{\vector( 0,-1){  6}}
\multiput(210,768)(30,0){2}{\vector( 0,-1){  6}}
\put(168,690){\vector( 1, 0){ 12}}
\multiput( 90,729)(30,0){3}{\vector( 0, 1){  6}}
\multiput( 90,705)(30,0){3}{\vector( 0,-1){  6}}
\multiput(210,708)(30,0){2}{\vector( 0,-1){  9}}
\multiput(210,729)(30,0){2}{\vector( 0, 1){  9}}
\put(258,738){\vector( 1, 1){0}}
\put(255,705){\vector( 1,-1){0}}
\put(258,762){\vector( 1,-1){0}}
\put(258,798){\vector( 1, 1){0}}
\put( 72,798){\vector(-1, 1){0}}
\put( 75,765){\vector(-1,-1){0}}
\put( 72,702){\vector(-1,-1){0}}
\put( 72,738){\vector(-1, 1){0}}
\put( 96,789){\scr$u\!\!+\!\!\lam$}
\put(102,732){\scr$u$}
\put(216,789){\scr$u\!\!+\!\!\lam$}
\put(216,759){\scr$u\!\!+\!\!\lam$}
\put( 96,759){\scr$u\!\!+\!\!\lam$}
\put(222,732){\scr$u$}
\put(222,702){\scr$u$}
\put(102,702){\scr$u$}
\put(246,777){\scr$u\!\!+\!\!\lam$}
\put( 63,777){\scr$u\!\!+\!\!\lam$}
\put( 69,717){\scr$u$}
\put(255,717){\scr$u$}
\end{picture}\ee
Inserting the crossing unitary condition (\ref{Wcrossunit})
into above term we have
\be
\setlength{\unitlength}{0.009in}%
\begin{picture}(280,120)(26,690)
\put(300,750){\circle*{4}}
\put(300,810){\line( 0,-1){120}}
\put(288,738){\vector( 1, 1){0}}
\multiput(285,705)(-30,30){2}{\vector( 1,-1){0}}
\put(288,762){\vector( 1,-1){0}}
\multiput(288,798)(-30,-30){2}{\vector( 1, 1){0}}
\put(300,810){\line(-1,-1){ 30}}
\put(270,780){\line( 1,-1){ 30}}
\put(300,750){\line(-1,-1){ 30}}
\put(270,720){\line( 1,-1){ 30}}
\put(276,777){\scr$u\!\!+\!\!\lam$}
\put(285,717){\scr$u$}
\put(270,780){\line(-1,-1){ 30}}
\put(240,750){\line( 1,-1){ 30}}
\put(258,744){\scr$2\!u\!\!+\!\!\lam$}
\put( 30,750){\circle*{4}}
\put( 30,810){\line( 1,-1){ 30}}
\put( 60,780){\line(-1,-1){ 30}}
\put( 30,750){\line( 1,-1){ 30}}
\put( 60,720){\line(-1,-1){ 30}}
\put( 30,810){\line( 0,-1){120}}
\multiput( 42,798)(30,-30){2}{\vector(-1, 1){0}}
\put( 45,765){\vector(-1,-1){0}}
\multiput( 42,702)(30,30){2}{\vector(-1,-1){0}}
\put( 42,738){\vector(-1, 1){0}}
\put( 33,777){\scr$u\!\!+\!\!\lam$}
\put( 39,717){\scr$u$}
\put( 60,780){\line( 1,-1){ 30}}
\put( 90,750){\line(-1,-1){ 30}}
\put( 42,744){\scr$-\!2\!u\!\!-\!\!3\!\lam$}
\put( 90,750){\circle*{4}}
\put(120,750){\circle*{4}}
\put(150,750){\circle*{4}}
\put(210,750){\circle*{4}}
\put(240,750){\circle*{4}}
\put( 90,810){\line( 1, 0){150}}
\put(240,810){\line( 0,-1){ 60}}
\put(240,750){\line( 0,-1){ 60}}
\put(240,690){\line(-1, 0){150}}
\put( 90,690){\line( 0, 1){120}}
\put( 90,780){\line( 1, 0){150}}
\put( 90,750){\line( 1, 0){150}}
\put( 90,720){\line( 1, 0){150}}
\put(120,810){\line( 0,-1){120}}
\put(150,810){\line( 0,-1){120}}
\put(210,810){\line( 0,-1){120}}
\put(168,690){\vector( 1, 0){ 12}}
\multiput(210,708)(30,0){2}{\vector( 0,-1){  9}}
\multiput( 90,708)(30,0){3}{\vector( 0,-1){  6}}
\multiput( 90,792)(30,0){3}{\vector( 0, 1){  6}}
\multiput( 90,738)(30,0){3}{\vector( 0,-1){  6}}
\multiput( 90,765)(30,0){3}{\vector( 0, 1){  6}}
\multiput(210,762)(30,0){2}{\vector( 0, 1){  9}}
\multiput(210,738)(30,0){2}{\vector( 0,-1){  6}}
\multiput(210,789)(30,0){2}{\vector( 0, 1){  6}}
\put( 96,789){\scr$u\!\!+\!\!\lam$}
\put(216,789){\scr$u\!\!+\!\!\lam$}
\put(222,702){\scr$u$}
\put(102,702){\scr$u$}
\put(102,762){\scr$u$}
\put( 96,729){\scr$u\!\!+\!\!\lam$}
\put(216,729){\scr$u\!\!+\!\!\lam$}
\put(225,762){\scr$u$}
\put( 63,780){\scr$a$}
\put( 82,735){\scr$d$}
\put( 33,807){\scr$c$}
\put( 84,804){\scr$c$}
\put( 84,774){\scr$a$}%b
\end{picture}\label{C2}
\ee
where we have not written down the ratio of the functions $S$,
which come from the crossing unitary condition and we have to
take care of them.
The sum over $a$ can be divided into the symmetric and antisymmetric
ones for $c=d$,
\be
&&\hs{-1}\sum_{a}{c,a,c\over{c,a,c}}=
{\mbox{$c,c-1,c$}\over{c,c-1,c}}+{\mbox{$c,c+1,c$}\over{c,c+1,c}}\no \\
&&=\(\sqrt{S(c-1)\over S(c+1)}{c,c-1,c\over c,c+1,c}
  +{c,c+1,c\over{c,c+1,c}}\)\no \\
&&\;\;\oplus\({c,c-1,c\over{c,c-1,c}}-
\sqrt{S(c-1)\over S(c+1)}{c,c-1,c\over c,c+1,c}\)\no
\ee
Therefore (\ref{C2}) is divided into two terms, which are
\be
\setlength{\unitlength}{0.009in}%
\begin{picture}(280,120)(-20,690)
\put(-80,744){$\disp{\sum_{b}}g_s(c,b,d)$}
\put(300,750){\circle*{4}}
\put(300,810){\line( 0,-1){120}}
\put(288,738){\vector( 1, 1){0}}
\multiput(285,705)(-30,30){2}{\vector( 1,-1){0}}
\put(288,762){\vector( 1,-1){0}}
\multiput(288,798)(-30,-30){2}{\vector( 1, 1){0}}
\put(300,810){\line(-1,-1){ 30}}
\put(270,780){\line( 1,-1){ 30}}
\put(300,750){\line(-1,-1){ 30}}
\put(270,720){\line( 1,-1){ 30}}
\put(276,777){\scr$u\!\!+\!\!\lam$}
\put(285,717){\scr$u$}
\put(270,780){\line(-1,-1){ 30}}
\put(240,750){\line( 1,-1){ 30}}
\put(258,744){\scr$2\!u\!\!+\!\!\lam$}
\put( 30,750){\circle*{4}}
\put( 30,810){\line( 1,-1){ 30}}
\put( 60,780){\line(-1,-1){ 30}}
\put( 30,750){\line( 1,-1){ 30}}
\put( 60,720){\line(-1,-1){ 30}}
\put( 30,810){\line( 0,-1){120}}
\multiput( 42,798)(30,-30){2}{\vector(-1, 1){0}}
\put( 45,765){\vector(-1,-1){0}}
\multiput( 42,702)(30,30){2}{\vector(-1,-1){0}}
\put( 42,738){\vector(-1, 1){0}}
\put( 33,777){\scr$u\!\!+\!\!\lam$}
\put( 39,717){\scr$u$}
\put( 60,780){\line( 1,-1){ 30}}
\put( 90,750){\line(-1,-1){ 30}}
\put( 42,744){\scr$-\!2\!u\!\!-\!\!3\!\lam$}
\put( 90,750){\circle*{4}}
\put(120,750){\circle*{4}}
\put(150,750){\circle*{4}}
\put(210,750){\circle*{4}}
\put(240,750){\circle*{4}}
\put( 90,810){\line( 1, 0){150}}
\put(240,810){\line( 0,-1){ 60}}
\put(240,750){\line( 0,-1){ 60}}
\put(240,690){\line(-1, 0){150}}
\put( 90,690){\line( 0, 1){120}}
\put( 90,780){\line( 1, 0){150}}
\put( 90,750){\line( 1, 0){150}}
\put( 90,720){\line( 1, 0){150}}
\put(120,810){\line( 0,-1){120}}
\put(150,810){\line( 0,-1){120}}
\put(210,810){\line( 0,-1){120}}
\put(168,690){\vector( 1, 0){ 12}}
\multiput(210,708)(30,0){2}{\vector( 0,-1){  9}}
\multiput( 90,708)(30,0){3}{\vector( 0,-1){  6}}
\multiput( 90,792)(30,0){3}{\vector( 0, 1){  6}}
\multiput( 90,738)(30,0){3}{\vector( 0,-1){  6}}
\multiput( 90,765)(30,0){3}{\vector( 0, 1){  6}}
\multiput(210,762)(30,0){2}{\vector( 0, 1){  9}}
\multiput(210,738)(30,0){2}{\vector( 0,-1){  6}}
\multiput(210,789)(30,0){2}{\vector( 0, 1){  6}}
\put( 96,789){\scr$u\!\!+\!\!\lam$}
\put(216,789){\scr$u\!\!+\!\!\lam$}
\put(222,702){\scr$u$}
\put(102,702){\scr$u$}
\put(102,762){\scr$u$}
\put( 96,729){\scr$u\!\!+\!\!\lam$}
\put(216,729){\scr$u\!\!+\!\!\lam$}
\put(225,762){\scr$u$}
\put( 63,780){\scr$a$}
\put( 82,735){\scr$d$}
\put( 33,807){\scr$c$}
\put( 84,804){\scr$c$}
\put( 84,774){\scr$b$}
\end{picture}\label{C3}
\ee
where $a=c+1$ for $c=d$ and $a=(c+d)/2$ otherwise and
\be
\setlength{\unitlength}{0.009in}%
\begin{picture}(280,120)(-20,690)
\put(-80,744){$\disp{\sum_{a}}g_a(c,a,d)$}
\put(300,750){\circle*{4}}
\put(300,810){\line( 0,-1){120}}
\put(288,738){\vector( 1, 1){0}}
\multiput(285,705)(-30,30){2}{\vector( 1,-1){0}}
\put(288,762){\vector( 1,-1){0}}
\multiput(288,798)(-30,-30){2}{\vector( 1, 1){0}}
\put(300,810){\line(-1,-1){ 30}}
\put(270,780){\line( 1,-1){ 30}}
\put(300,750){\line(-1,-1){ 30}}
\put(270,720){\line( 1,-1){ 30}}
\put(276,777){\scr$u\!\!+\!\!\lam$}
\put(285,717){\scr$u$}
\put(270,780){\line(-1,-1){ 30}}
\put(240,750){\line( 1,-1){ 30}}
\put(258,744){\scr$2\!u\!\!+\!\!\lam$}
\put( 30,750){\circle*{4}}
\put( 30,810){\line( 1,-1){ 30}}
\put( 60,780){\line(-1,-1){ 30}}
\put( 30,750){\line( 1,-1){ 30}}
\put( 60,720){\line(-1,-1){ 30}}
\put( 30,810){\line( 0,-1){120}}
\multiput( 42,798)(30,-30){2}{\vector(-1, 1){0}}
\put( 45,765){\vector(-1,-1){0}}
\multiput( 42,702)(30,30){2}{\vector(-1,-1){0}}
\put( 42,738){\vector(-1, 1){0}}
\put( 33,777){\scr$u\!\!+\!\!\lam$}
\put( 39,717){\scr$u$}
\put( 60,780){\line( 1,-1){ 30}}
\put( 90,750){\line(-1,-1){ 30}}
\put( 42,744){\scr$-\!2\!u\!\!-\!\!3\!\lam$}
\put( 90,750){\circle*{4}}
\put(120,750){\circle*{4}}
\put(150,750){\circle*{4}}
\put(210,750){\circle*{4}}
\put(240,750){\circle*{4}}
\put( 90,810){\line( 1, 0){150}}
\put(240,810){\line( 0,-1){ 60}}
\put(240,750){\line( 0,-1){ 60}}
\put(240,690){\line(-1, 0){150}}
\put( 90,690){\line( 0, 1){120}}
\put( 90,780){\line( 1, 0){150}}
\put( 90,750){\line( 1, 0){150}}
\put( 90,720){\line( 1, 0){150}}
\put(120,810){\line( 0,-1){120}}
\put(150,810){\line( 0,-1){120}}
\put(210,810){\line( 0,-1){120}}
\put(168,690){\vector( 1, 0){ 12}}
\multiput(210,708)(30,0){2}{\vector( 0,-1){  9}}
\multiput( 90,708)(30,0){3}{\vector( 0,-1){  6}}
\multiput( 90,792)(30,0){3}{\vector( 0, 1){  6}}
\multiput( 90,738)(30,0){3}{\vector( 0,-1){  6}}
\multiput( 90,765)(30,0){3}{\vector( 0, 1){  6}}
\multiput(210,762)(30,0){2}{\vector( 0, 1){  9}}
\multiput(210,738)(30,0){2}{\vector( 0,-1){  6}}
\multiput(210,789)(30,0){2}{\vector( 0, 1){  6}}
\put( 96,789){\scr$u\!\!+\!\!\lam$}
\put(216,789){\scr$u\!\!+\!\!\lam$}
\put(222,702){\scr$u$}
\put(102,702){\scr$u$}
\put(102,762){\scr$u$}
\put( 96,729){\scr$u\!\!+\!\!\lam$}
\put(216,729){\scr$u\!\!+\!\!\lam$}
\put(225,762){\scr$u$}
\put( 63,780){\scr$a$}
\put( 82,735){\scr$d$}
\put( 33,807){\scr$c$}
\put( 84,804){\scr$c$}
\put( 84,774){\scr$b$}
\end{picture}\label{C4}
\ee
where $c=d$ and $b=c-1$. The factors $g_s$ and $g_a$ are given by
\be
g_s(a,b,c)&=&\cases{\disp{\sqrt{S(a-1)\over S(a+1)}},
  & $b=a-1,a=c$ \cr       \h 1,     & otherwise}      \\
g_a(a,b,c)&=&\cases{-\disp{\sqrt{S(a-1)\over S(a+1)}},
  & $b=a+1,a=c$ \cr     \h \delta_{a,c},     & otherwise}
\ee
The first term is the fused transfer matrix of the
model at level $2$, which
is expressed by $\V^{(2)}(u)=\V^{(2)}_0$. The second
term is the antisymmetric
fusion of the model, which gives exactly the factor
$f^{1}_0=f^{1}(u)$. Note
that the function $f^{1}(u)$ is dependent the face
boundary $K_\pm$ matrices.
So we finally have the functional equations (\ref{func-V}).


\begin{thebibliography}{99}
\bibitem{comm0} In \cite{PeBe:95} the reflection equations do not
   follow from the orignal formulation given by Sklyanin in
   \cite{Sklyanin} and thus they are set up differently.
\end{thebibliography}
\end{document}